 \documentclass[twocolumn]{aastex6}

\fullcollaborationName{The Friends of AASTeX Collaboration}
\begin{document}


\title{ALMA observations and modeling of the rotating outflow in Orion Source I}



\author{J.A. L\'opez-V\'azquez\altaffilmark{1}, Luis A. Zapata\altaffilmark{1}, Susana Lizano\altaffilmark{1}, and Jorge Cant\'o\altaffilmark{2}} 


\altaffiltext{1}{Instituto de Radioastronom\'ia y Astrof\'isica, Universidad Nacional Aut\'onoma de M\'exico, \\ Apartado Postal 3-72, 58089 Morelia, Michoac\'an, M\'exico}
\altaffiltext{2}{Instituto de Astronom\'ia, Universidad Nacional Aut\'onoma de M\'exico,\\ Apartado Postal 70-264, 04510, CDMX, M\'exico}


\begin{abstract}
We present $^{29}$SiO(J=8--7) $\nu$=0, SiS (J=19--18) $\nu$=0, and $^{28}$SiO (J=8--7) $\nu$=1 molecular line archive
observations made with the Atacama Large Millimeter/Submillimeter Array (ALMA) of the molecular outflow 
associated with Orion Source I. The observations show velocity asymmetries about the flow axis which are 
interpreted as outflow rotation. We find that the rotation velocity ($\sim$4--8 km s$^{-1}$)
 decreases with the vertical distance to the disk. In contrast,
the cylindrical radius ($\sim$100--300 au), the expansion velocity ($\sim$2--15 km s$^{-1}$), and the axial velocity $v_{\rm z}$ ($\sim$-1--10 km s$^{-1}$)
increase with the vertical distance. The mass estimated of the molecular outflow $\mathrm{M}_{\rm outflow}\sim$0.66--1.3 M$_\odot$.
Given a kinematic time $\sim$130 yr, this implies a mass loss rate $\dot{\mathrm{M}}_{\rm outflow} \sim 5.1-10 \times 10^{-3}$ M$_\odot$ yr$^{-1}$.
This massive outflow sets important contraints on disk wind models.  We 
compare the observations with a model of a shell produced by the interaction between an anisotropic stellar wind and an Ulrich accretion flow that 
corresponds to a rotating molecular envelope in collapse. We find that the model cylindrical radii are 
consistent with the $^{29}$SiO(J=8--7) $\nu$=0 data.
The expansion velocities and the axial velocities of the model are similar the observed values, except close to the disk ($z\sim\pm$150 au) for the expansion velocity. Nevertheless, the rotation velocities of the model 
are a factor $\sim$3--10 lower 
 than the observed values.
 We conclude that the Ulrich flow alone cannot explain the rotation observed
and other possibilities should be explored, like the inclusion of the angular momentum of a disk wind.
\end{abstract}

\keywords{accretion -- ISM: jets and outflows -- stars: individual (Orion Source I, Kleinmann-Low Nebula) -- pre-main sequence}

\section{Introduction}
\label{sec:introduction}

The molecular outflows and the protostellar jets are present in the star formation process and appear to be more powerful and collimated during the earliest phases of young stellar sources (e.g., \citealt{Bontemps_1996}), however, their origin is under debate. Two scenarios are proposed to explain the formation of the molecular outflows. In the first case, several authors (e.g., \citealt{Pudritz_1986}, \citealt{Launhardt_2009}, and \citealt{Pech_2012}), propose that the molecular outflows are ejected directly from the accretion disk. Other authors suggest (e.g., \citealt{Shu_1991}, \citealt{Canto_1991}, and \citealt{Ragacabrit_1993}), that the molecular outflows are a mixture between the entrained material with from the molecular cloud and a fast stellar wind.

The magnetocentrifugal mechanism \citep{Blandford_1982} is the principal candidate
 for producing jets and stellar winds (see reviews by \citealt{Konigl_2000} and \citealt{Shu_2000}), in this mechanism, the rotating magnetic field anchored to the star--disk system drives and collimates these winds \citep{Pudritz_2007,Shang_2007}.
Nevertheless, it is not clear where the magnetic
 fields are anchored to the disk. The magnetocentrifugal mechanism has two different origins: X-wind (\citealt{Shu_1994}) and disk winds (\citealt{Pudritz_1983}). 
 In the first model, these winds are launched close to the star, from the radius where the stellar magnetosphere truncates the disk. In the second model, 
 these winds come from a wider range of the radii. \citet{Anderson_2003} found a general relation between the poloidal and toroidal velocity 
 components of the magneto-centrifugal winds at large distances and the rotation velocity at the ejection point. Therefore, 
 the observed rotation velocity of the jet could give information about its origin on the disk.

In recent years, evidence of the rotation in protostellar jets and the molecular outflows has been found. For example, the jets HH 211 (\citealt{Lee_2009}) and HH 212 (\citealt{Lee_2017}) present signature of the rotation of a few km s$^{-1}$. Molecular outflows with signature of the rotation are: CB 26 (\citealt{Launhardt_2009}), Ori-S6 \citep{Zapata_2010}, HH 797 (\citealt{Pech_2012}), DG Tau B (\citealt{Zapata_2015}), Orion Source I \citep{Hirota_2017}, HH 212 \citep{Tabone_2017}, HH 30 \citep{Louvet_2018}, and NGC 1333 IRAS 4C \citep{Zhang_2018}.

\begin{figure*}[ht]
\centering
\includegraphics[scale=0.34]{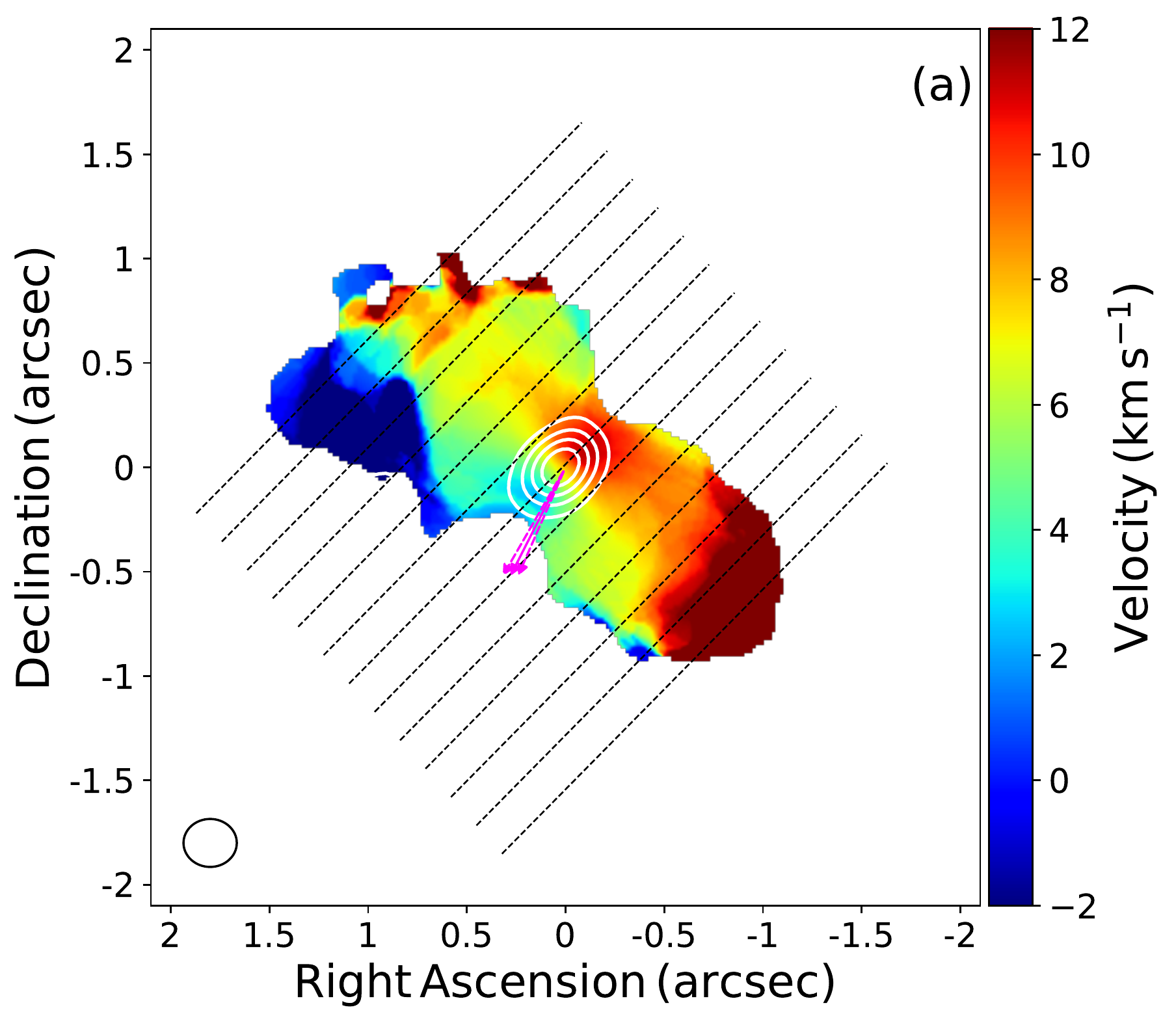}\includegraphics[scale=0.34]{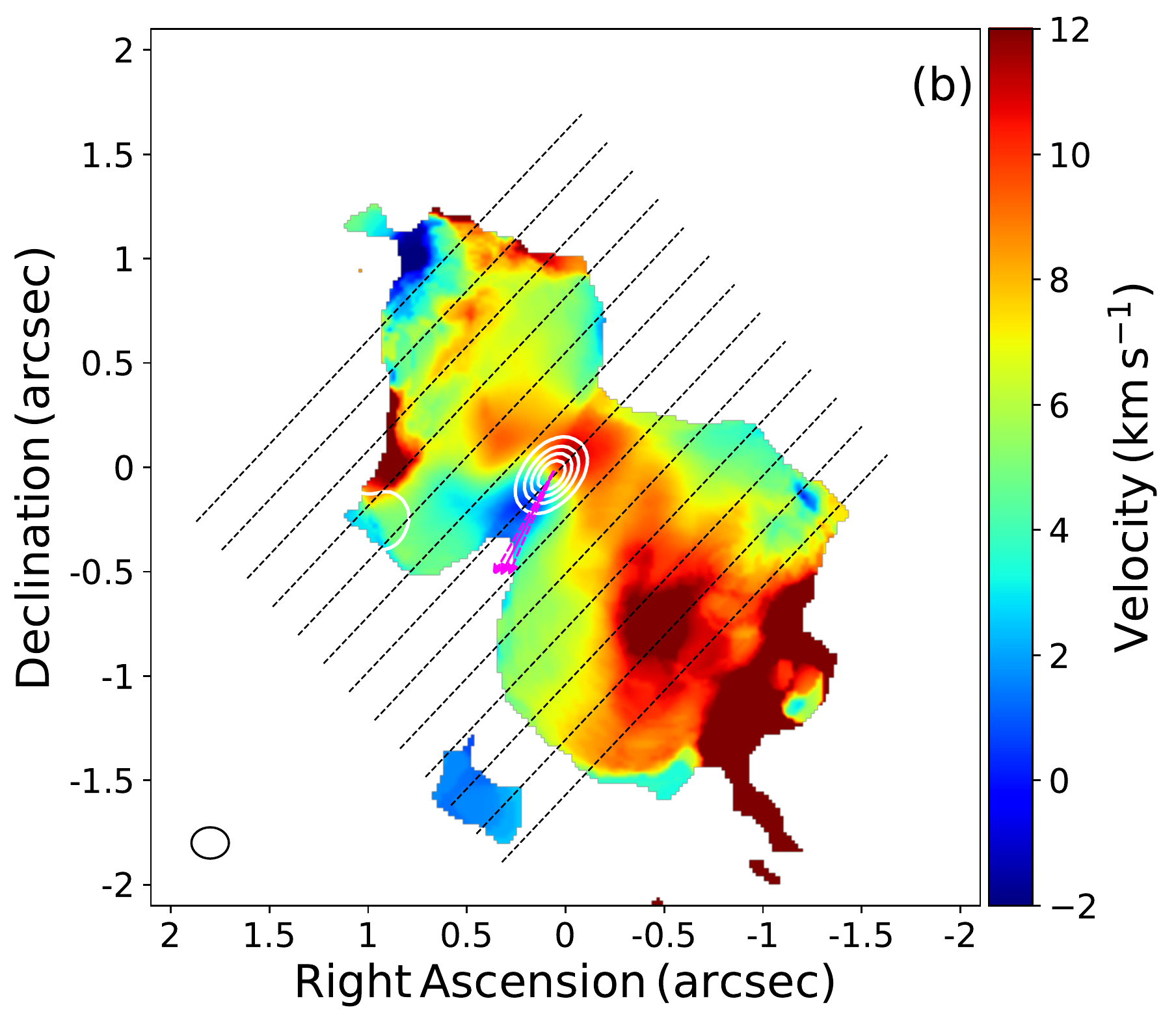}\includegraphics[scale=0.34]{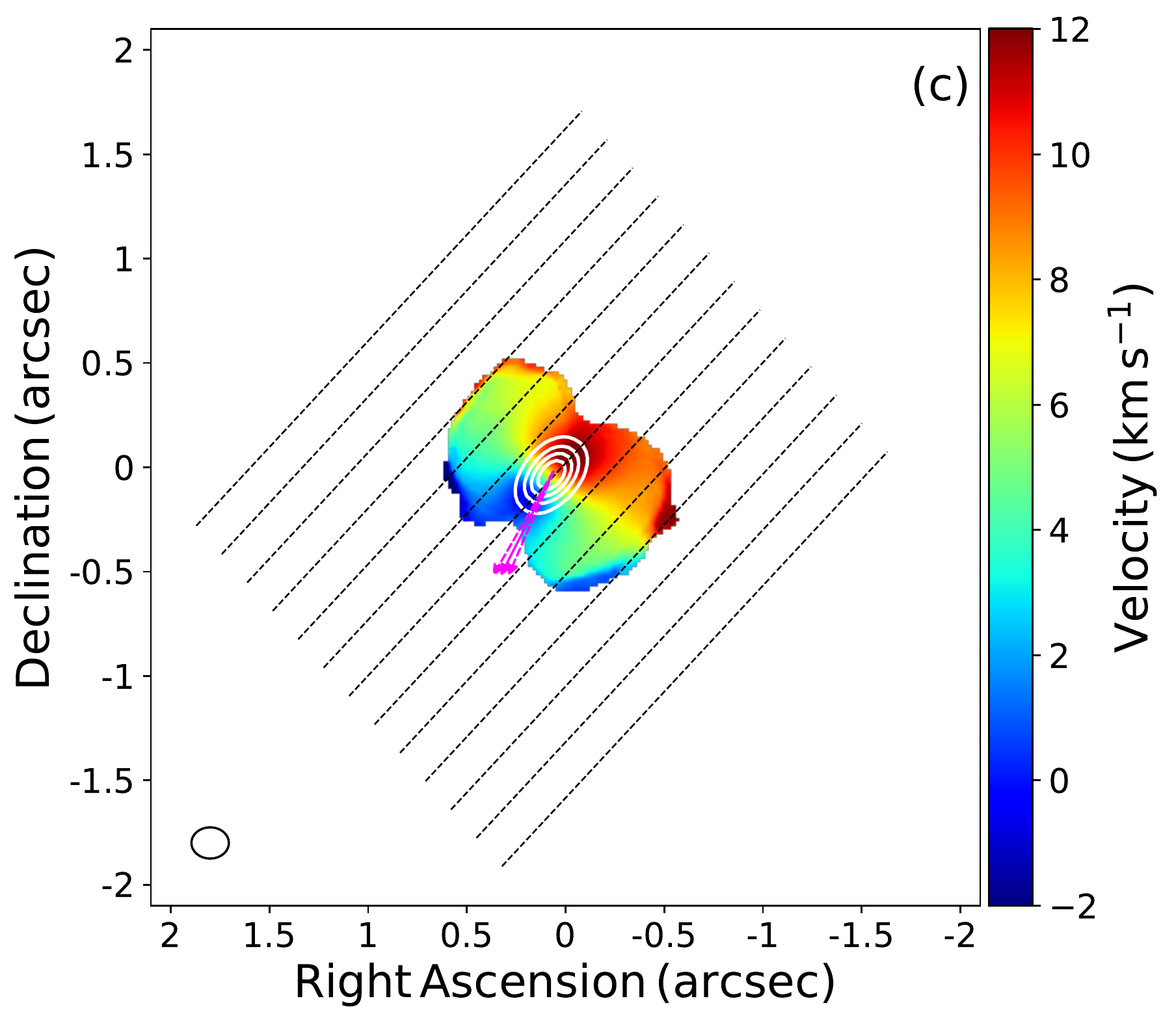}
\caption{\scriptsize ALMA first moment or the intensity weighted velocity of the emission from the different molecule lines from the outflow. 
(a) Emission of $^{29}$SiO (J=8--7) $\nu$=0. (b) Emission from SiS (J=19--18) $\nu$=0. 
(c) Emission from $^{28}$SiO (J=8--7) $\nu$=1. The diagonal dashed lines indicate the positions where the position-velocity diagrams were made. 
The color scale bar on the right side shows the $V_{\rm LSR}$ in km s$^{-1}$. The synthesized beam of the image is shown in the lower left corner. 
In these plots the white contours show the continuum emission from the disk and are the 20$\sigma$, 40$\sigma$, 60$\sigma$, and 80$\sigma$. The magenta arrows indicate the proper motion of this source. The size these arrows indicates the proper motions for a period of 100 years. The solid arrows indicate the proper motion and the dashed arrows indicate the proper motion considering the error in right ascension.}
\label{fig:mom1-29sio}
\end{figure*}

\citet{Zapata_2015}, argued that slow winds ejected from large disk radii do not have enough mass, thus, these winds cannot account for the observed linear and angular momentum rates of the molecular outflow of DG Tau B. 
Their argument assumed that the mass loss rate of the wind is a small fraction $f \sim 0.1$ of the disk mass accretion rate ($\dot M_w \sim f \dot M_{\rm d,a}$). Nevertheless, recent 
non-ideal magnetohydrodynamic simulations of magnetized disk winds 
show that this fraction can be very large, $f \sim 1 - 2 $ (e.g., \citealt{Bai_2017}; \citealt{Wang_2019}).
However,  massive disk winds could pose a problem to the  disk lifetime. The mass of the disk of  DG Tau B  is $M_d \sim 0.068 M_\odot$ (\citealt{Guilloteau_2011}). Given the observed outflow mass loss rate 
$1.7 -2.9 \times 10^{-7} M_\odot {\rm yr}^{-1}$ (\citealt{Valon_2020}), 
the disk lifetime is 
$\tau = M_d / \dot M_{\rm d, a}  = f M_d / \dot M_{\rm w}  \sim f (2 - 4 ) \times 10^5 {\rm \, yr}$.  Depending on the value of $f$, the disk lifetime could be smaller than the age of DG Tau B,
which  has been cataloged as a Class I/II source (\citealt{Hartmann_2005}; \citealt{Luhman_2010}). 

The large masses of the molecular outflows can be explained if the outflow is formed mainly by entrained material from the parent cloud. 
\citet{JALV2019}, hereafter LV19, modeled the molecular outflow as a thin shocked shell, formed by the collision between an anisotropic 
stellar wind and a rotating molecular cloud in collapse, described by \citet{Ulrich_1976}. They found that the mass of the molecular outflow, 
probably, comes from the parent cloud, but the angular momentum could come from both the stellar wind and the parent cloud. 

Located at the center of the Kleinmann-Low Nebula in Orion, at a distance $\sim$ 418$\pm$ 6 pc (\citealt{Kim_2008}), the Orion Source I (Orion  Src I) is 
a candidate high mass (M$_*>$ 8 M$_{\odot}$) star (\citealt{Hirota_2014}; \citealt{Plambeck_2016}; \citealt{Hirota_2017}; \citealt{Ginsburg_2018}). 
The central object of the Orion Src I has a high luminosity $\sim$ 10$^{4}$ L$_\odot$ (\citealt{Menten_1995}; \citealt{Reid_2007}; \citealt{Testi_2010}). 
The bipolar outflow presents low radial velocities ($\sim$ 18 km s$^{-1}$) along the northeast-southwest direction, 
with a size $\sim$ 1000 au (\citealt{Plambeck_2009}; \citealt{Zapata_2012}; \citealt{Greenhill_2013}). This source has a proper motion respect to the nebula center of $\mu_\alpha \cos\delta=+2.9\pm0.4$ mas yr$^{-1}$ and $\mu_\delta=-5.4\pm0.4$ mas yr$^{-1}$, where the angle $\delta\sim -5^\circ$ (\citealt{Rodriguez_2017}).
In fact, the Orion Kleinmann-Low Nebula exhibits evidence of a violent explosive phenomenon (e.g., \citealt{Bally_2005}; \citealt{Gomez_2008}; \citealt{Zapata_2009}; \citealt{Bally_2017}; \citealt{Zapata_2017}). The proper motions of the sources I, BN, and n reveal that this explosion appears to have taken place 500 year ago (e.g., \citealt{Luhman_2017}; \citealt{Rodriguez_2017}).

We present archive $^{29}$SiO (J=8--7) $\nu$=0, SiS (J=19--18) $\nu$=0, and $^{28}$SiO (J=8--7) $\nu$=1 line observation, made with the Atacama 
Large Millimeter/Submillimeter Array (ALMA) of the molecular outflow associated with the young star Orion Src I. We also compare 
the observational results with the thin shell model of LV19. The paper is organized as follows: The Section \ref{sec:observations} details the observations. In Section \ref{sec:results} we present our observational results and compare with the outflow model. Finally, the conclusions are presented in Section \ref{sec:Conclusions}.

\begin{table}[!t]
 \centering
 \setlength\tabcolsep{2.0pt}
\caption{Molecular lines.}
\begin{tabular}{c c}
 \hline
 \hline
\textbf{Molecular}  & \textbf{Rest } \\
 \textbf{Specie   }   & \textbf{Frequency}\\
 & [GHz]\\
\hline
$^{29}$SiO(J=8--7)  $\nu=0$ & 342.9808 \\
SiS (J=19--18) $\nu=0$ & 344.7794 \\
$^{28}$SiO(J=8--7) $\nu=1$ & 344.9162\\
\hline
\end{tabular}
  \label{tab:molecules}
\end{table}

\begin{figure*}[h!]
\centering
\includegraphics[scale=0.255]{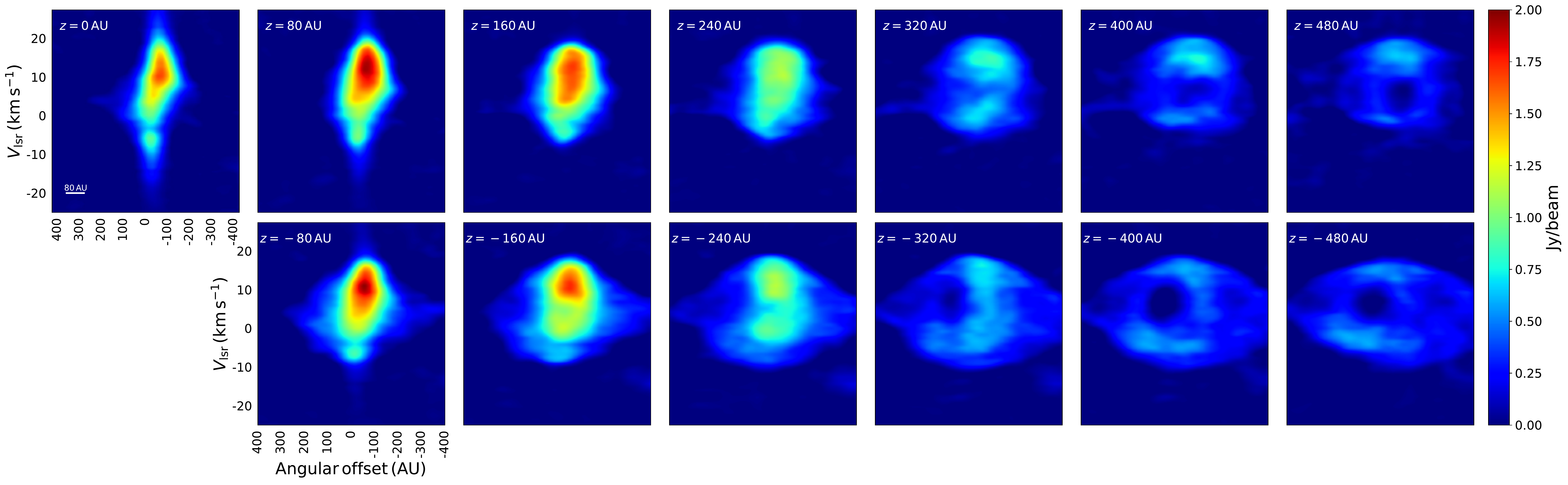}
\caption{\scriptsize Position-velocity diagrams parallel to the disk mid-plane from the emission of the 
$^{29}$SiO (J=8--7) $\nu$=0 transition at different heights from $z=480$ au to $z=-480$ au with an interval of 80 au. 
The vertical axes are the line of sight velocity with respect to the LSR velocity and the horizontal axes are the perpendicular distances 
with respect to the outflow axis. The color scale bar on the right side shows the intensity in Jy/beam.}
\label{fig:pv29sio}
\end{figure*}

\begin{figure*}[h!]
\centering
\includegraphics[scale=0.255]{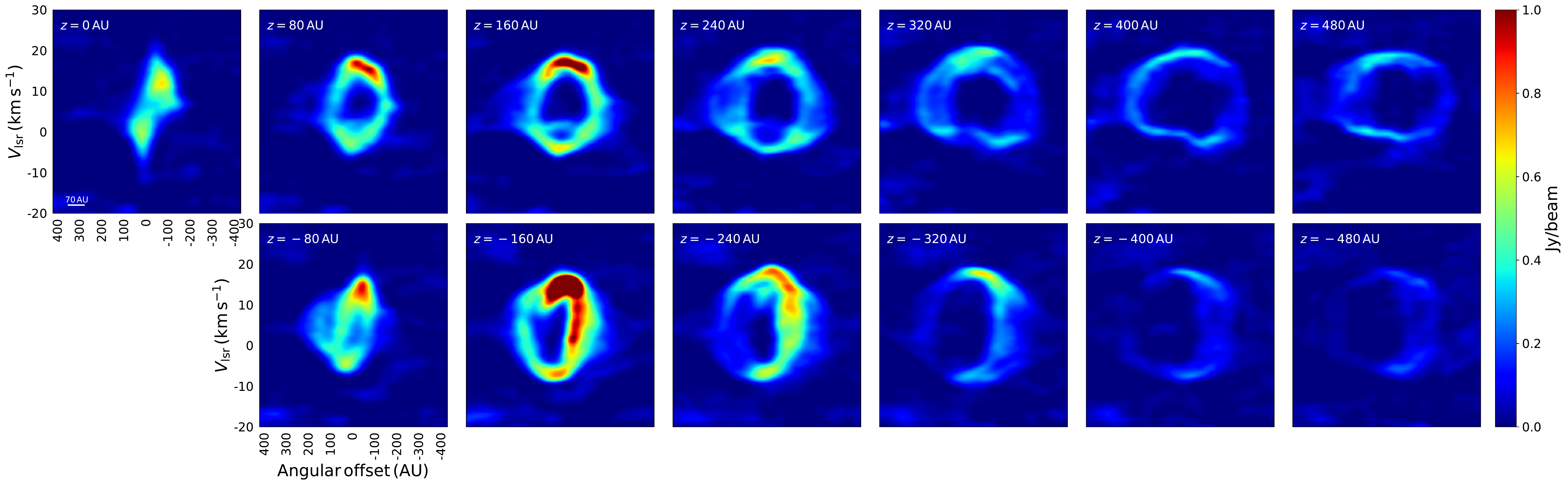}
\caption{\scriptsize Position-velocity diagrams parallel to the disk mid-plane from the emission of the SiS (J=19--18) $\nu$=0 transition
for the same heights and the same description as Figure \ref{fig:pv29sio}.}
\label{fig:pvsis}
\end{figure*}

\begin{figure*}[t!]
\centering
\includegraphics[scale=0.255]{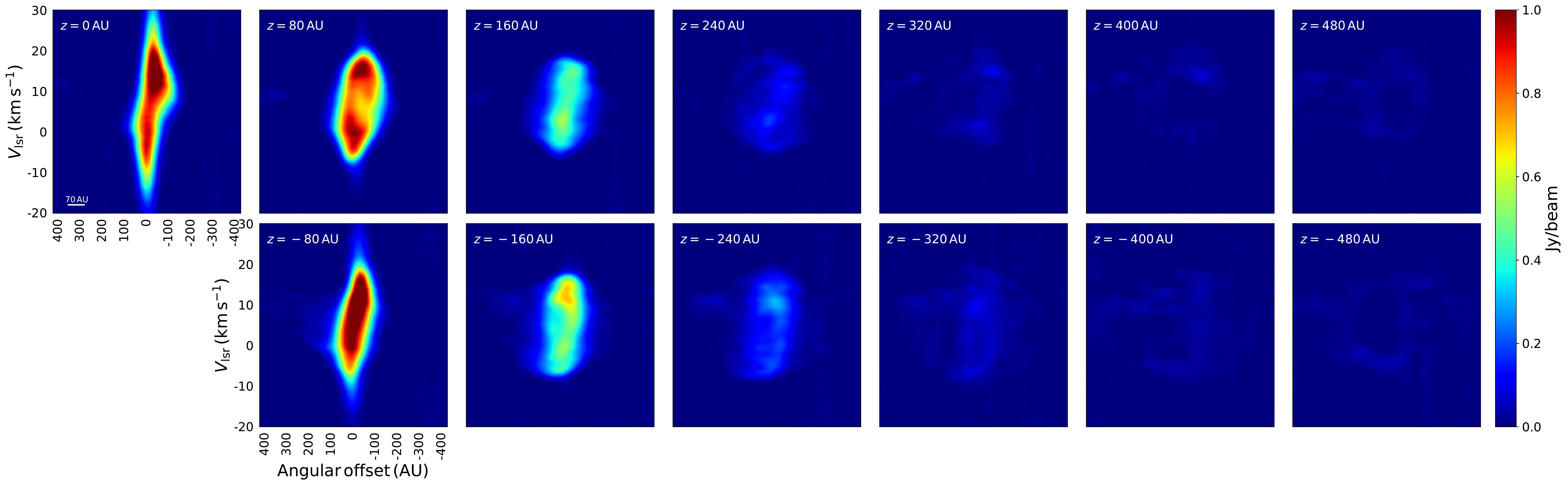}
\caption{\scriptsize Position-velocity diagrams parallel to the disk mid-plane from the emission of the $^{28}$SiO (J=8-7) $\nu$=1 transition
for the same heights and the same description as Figure \ref{fig:pv29sio}.}
\label{fig:pvsio}
\end{figure*}

\begin{figure*}[t!]
\centering
\includegraphics[scale=0.55]{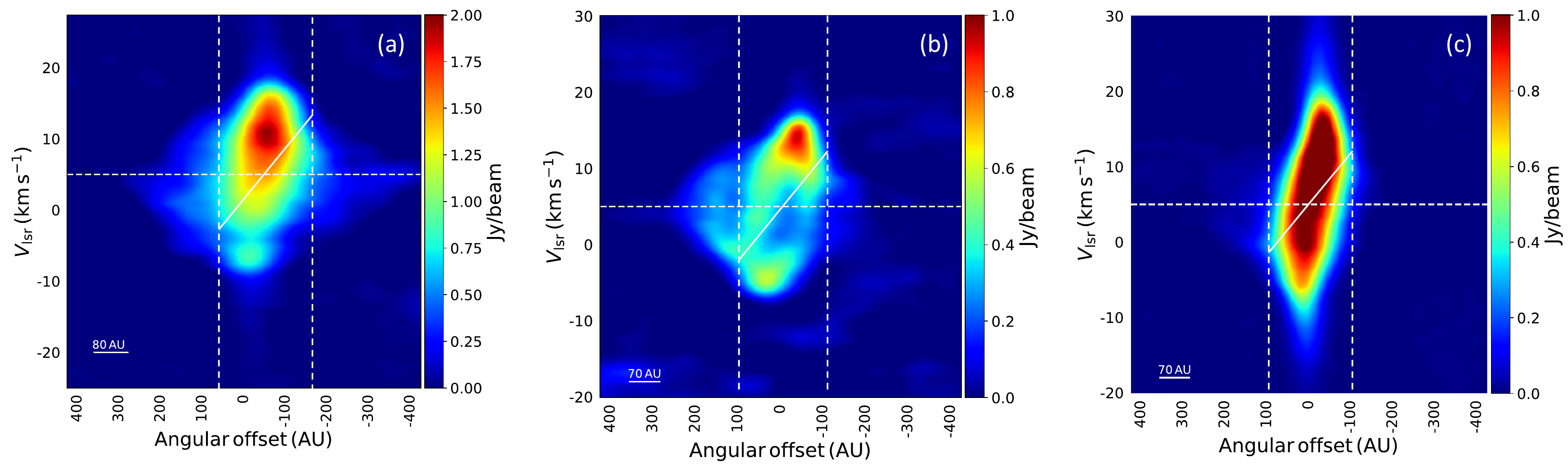}
\caption{\scriptsize Position-velocity diagrams parallel to the disk midplane at a height $z=-80$ au. 
(a) Emission from the molecular line of $^{29}$SiO (J=8--7) $\nu=0$. 
(b) Emission from the molecular line SiS (J=19--18) $\nu=0$. 
(c) Emission from the molecular line of $^{28}$SiO (J=8--7) $\nu=1$. 
The horizontal dashed line shows the value of the LSR velocity of the source. 
The vertical dashed lines represent the cylindrical radius $\varpi_{\rm obs}$ defined in Figure \ref{fig:esquema}. The solid line in each panel indicates the rotation signature.}
\label{fig:pvrot}
\end{figure*}

\section{Observations}
\label{sec:observations}

The archive observations of Orion Src I were carried out with the  Atacama Large Millimeter/Submillimeter Array (ALMA) in band 7 in 2016 October 31st and
2014 July 26th as part of the programs 2016.1.00165.S (P.I. John Bally) and  2012.1.00123.S (P.I. Richard Plambeck), respectively. At that time, the array counted 
with 31 (2014) and 42 (2016) antennas with a diameter of 12m yielding baselines with projected lengths from 33 to 820 m (41 -- 1025 k$\lambda$) 
and 18 to 1100 m (22 -- 1375 k$\lambda$), respectively. The primary beam at this frequency has a full width at half-maximum (FWHM) of about 20$''$, so that
in both observations the molecular emission from the outflow of Orion Src I falls well inside of this area.  

The integration time on-source was about 25 min., and 32 min. was used for calibration for the 2014 observations, while for the 2016 observations was about 
13 min. on-source, and 37 min. for calibration. The ALMA digital correlator was configured with four spectral windows centered at 353.612 GHz (spw0), 
 355.482 GHz (spw1),  341.493 GHz (spw2), and 343.363 GHz (spw3) with 3840 channels and a space channel of 488.281 kHz or about 0.4 km s$^{-1}$ for the 2014 observations
 and at  344.990 GHz (spw0), 346.990 GHz (spw1),  334.882 GHz (spw2), and 332.990 GHz (spw3) with 1920 channels and a space channel of 976.562 kHz 
 or about 0.8 km s$^{-1}$ for the 2016 observations.  The spectral lines reported on this study were found in the spw2 ($^{29}$SiO) of the 2014 observations and
 the spw0 (SiO and SiS) of the 2016 observations (see Table \ref{tab:molecules}).

For both observations, the weather conditions were reasonably good and stable for these high frequencies. The observations used the quasars: 
J0510$+$1800, J0522$-$3627, J0527$+$0331, J0532$-$0307, J0607$-$0834, J0423$-$013 and J0541$-$0541 for amplitud, 
phase, bandpass, pointing, water vapor radiometer, and atmosphere calibration.

The data were calibrated, imaged, and analyzed using the Common Astronomy Software Applications (CASA Version 5.1). The resulting image rms noises for the spectral lines were
about 10 mJy Beam$^{-1}$ (SiO and SiS) at a angular resolution of 0.19$''$ $\times$ 0.14$''$ with a PA of $-$63$^\circ$ and about 20 mJy Beam$^{-1}$ ($^{29}$SiO) 
at an angular resolution of 0.30$''$ $\times$ 0.24$''$ with a PA of $+$58$^\circ$. Self-calibration was attempted on the continuum, however, we did not obtain a relatively good improvement
in the line maps.    

\begin{figure*}[t!]
\centering
\includegraphics[scale=0.55]{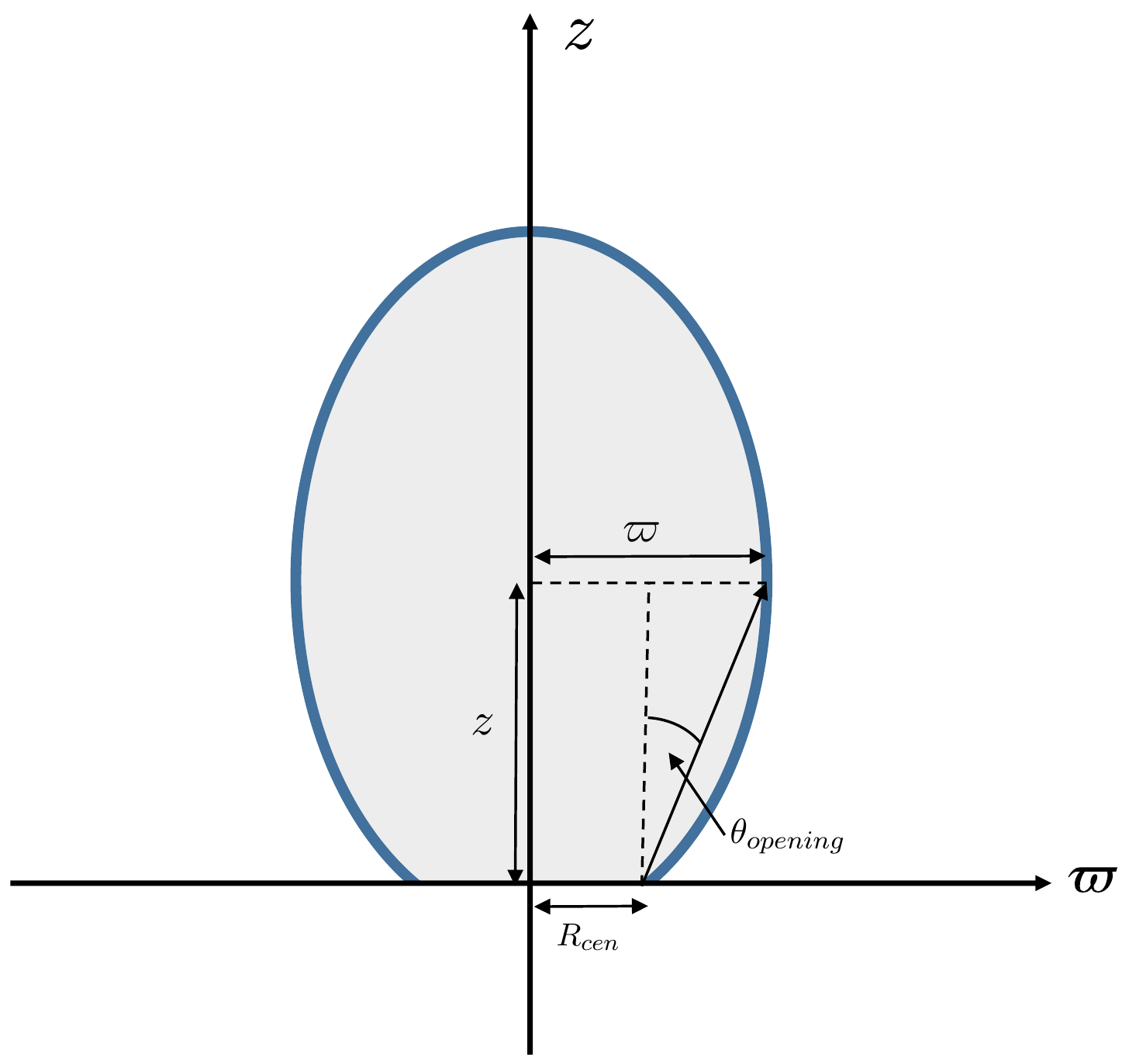}
\caption{Schematic diagram of a molecular outflow. 
This diagram shows the opening angle of the molecular outflow $\theta_{\rm opening}$, 
the cylindrical radius $\varpi$, the centrifugal radius $R_{\rm cen}$, and the height $z$.}
\label{fig:esquema}
\end{figure*}

\begin{figure*}[t!]
\centering
\includegraphics[scale=0.7]{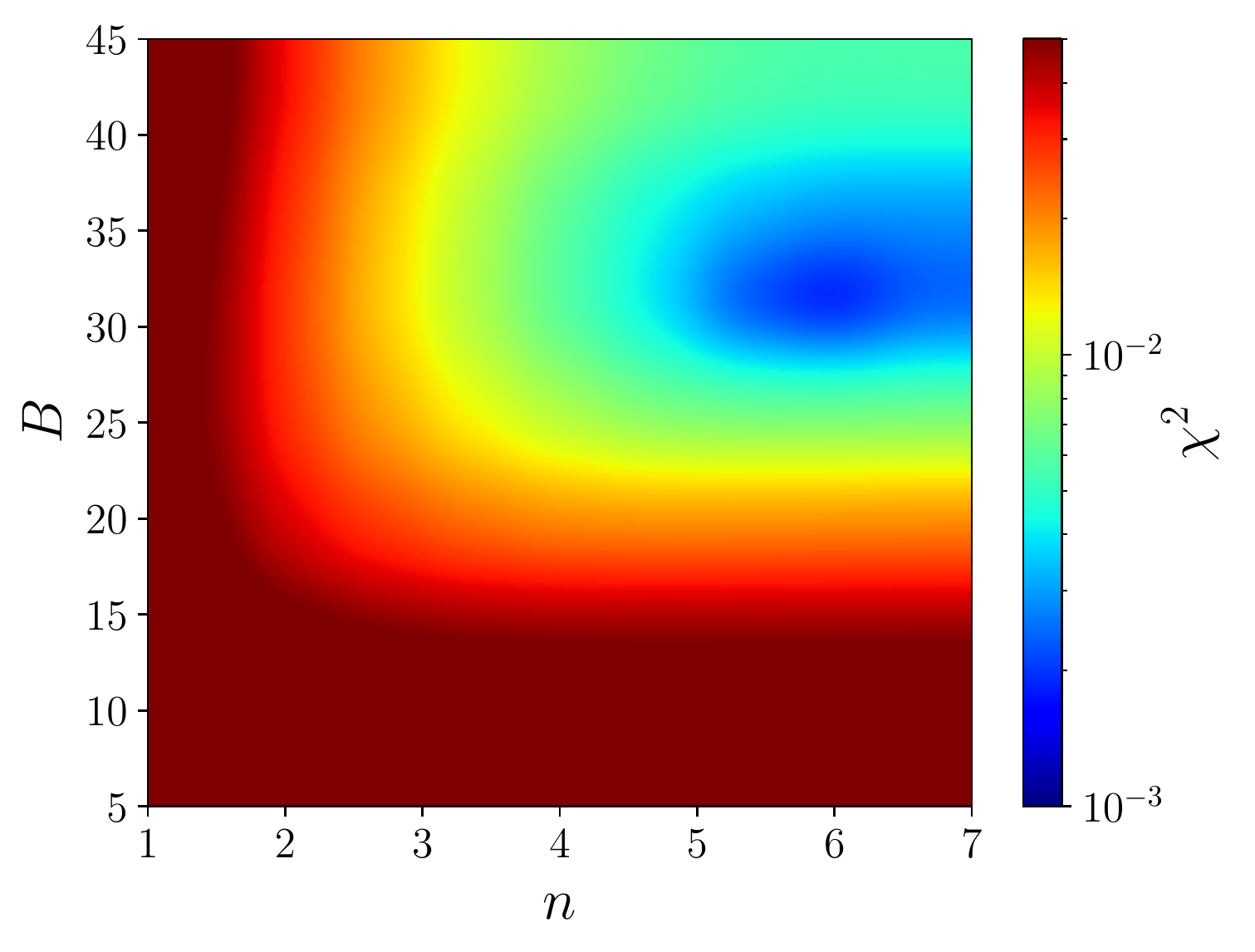}
\caption{$\chi^2$ analysis for different anisotropy parameters $B$ and $n$,  with $A=1$, $\alpha=0.1$, $\beta=4$, and an integration time of $t=65$ yr.}
\label{fig:chisq}
\end{figure*}

\begin{figure*}[t!]
\centering
\includegraphics[scale=0.34]{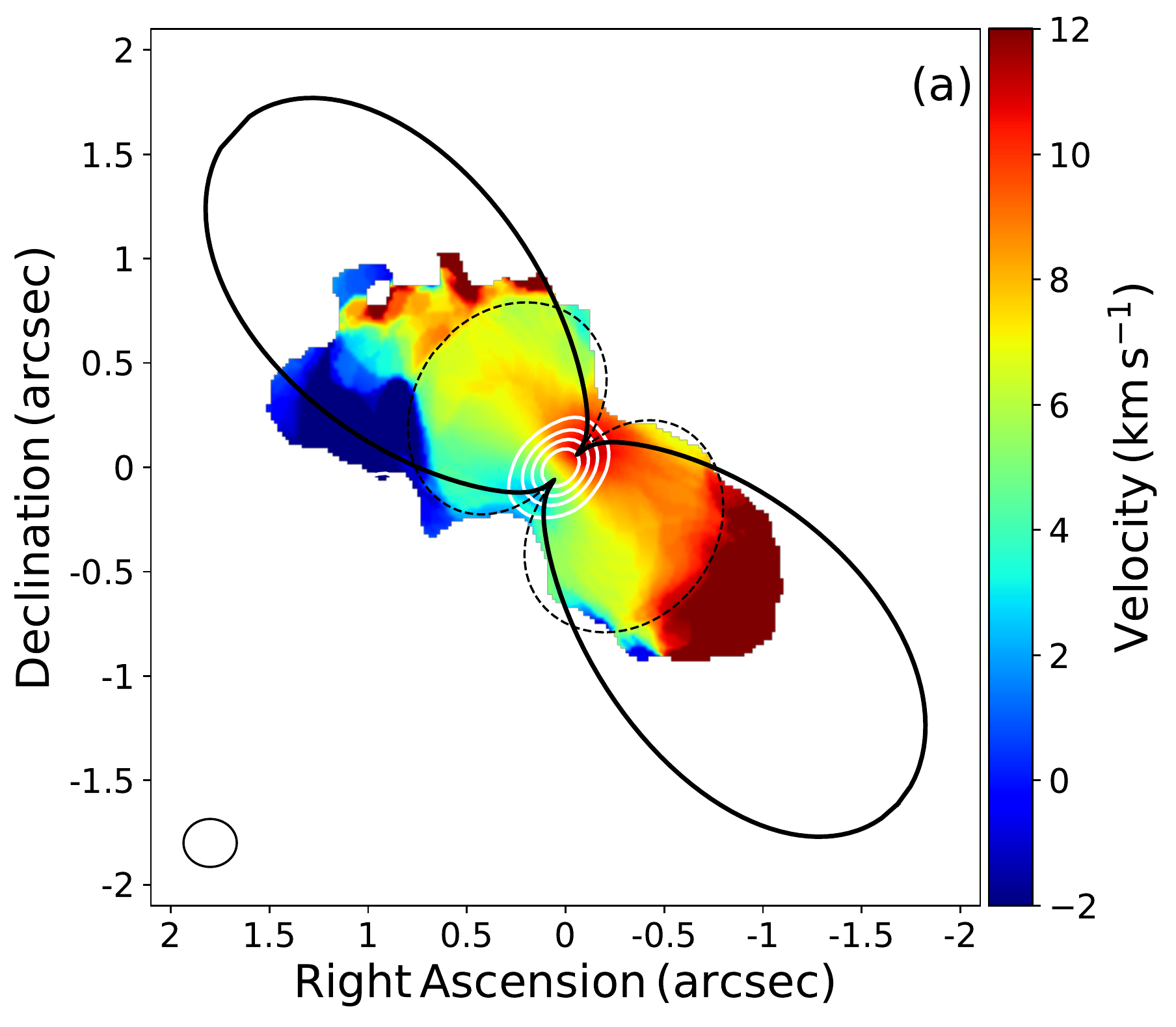}\includegraphics[scale=0.34]{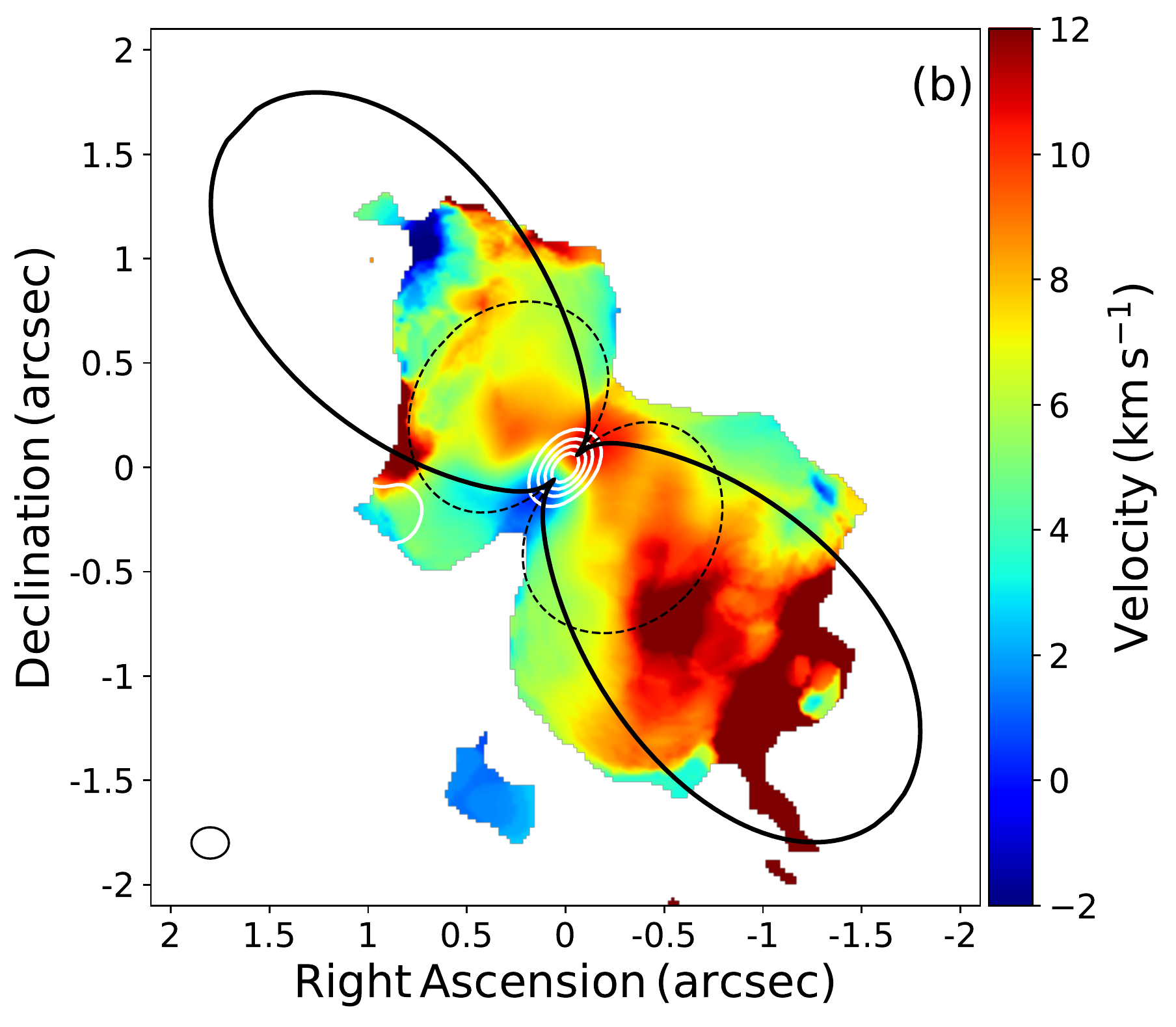}\includegraphics[scale=0.34]{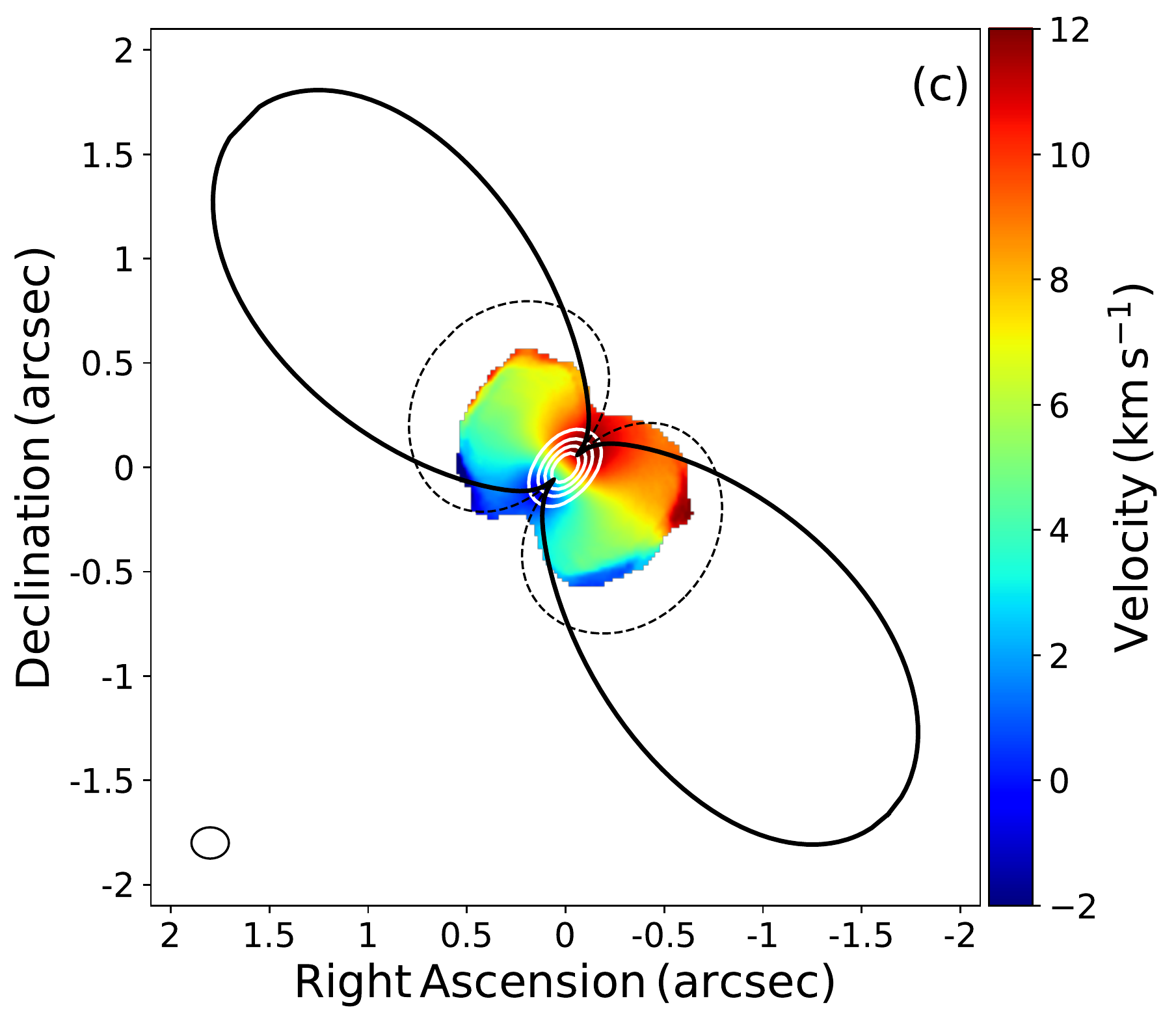}
\caption{Comparison between the ALMA first moment or the intensity weighted velocity of the emission from the different molecule lines with the best 
outflow model (see text). (a) Emission of $^{29}$SiO (J=8--7) $\nu$=0. (b) Emission from SiS (J=19--18) $\nu$=0. (c) Emission from $^{28}$SiO (J=8--7) $\nu$=1. The dashed line represents the isotropic model for parameters $\alpha=0.1$, $\beta=4$, and $B=0$. The solid black line represents the outflow model for the parameters $\alpha$=0.1, $\beta$=4, $r_{s0}(0)=10^{-3}$, $A$=1, $B$=35, and $n$=5.}
\label{fig:bestmodel}
\end{figure*}

\section{Results and Discussion}
\label{sec:results}

\citet{Hirota_2017} present observational results with ALMA  at 50 au resolution from the emission of the Si$^{18}$O and H$_2$O molecular lines of the molecular outflow of Orion Src I. These lines trace the inner part of the molecular outflow. In contrast, the archive observations used in this work trace the outer part of the molecular outflow, the latter, because this improves an easily comparison with the thin shell model of LV19.

\subsection{Results from the observations}
\label{subsec:observations}

Figure \ref{fig:mom1-29sio} presents the first moment or the intensity weighted velocity of the emission from the three molecular lines, $^{29}$SiO (J=8--7) $\nu$=0 (panel a), 
SiS (J=19--18) $\nu$=0 (panel b), and $^{28}$SiO (J=8--7) $\nu$=1 (panel c). These panels show that the east side of the molecular outflow presents blueshifted velocities, 
while the west side presents redshifted velocities. This difference of the velocity is interpreted as rotation around the outflow axis \citep{Hirota_2017}.  Moreover, Figure \ref{fig:mom1-29sio} indicates that the molecular 
outflow is not on the plane of the sky, i.e., the outflow has an inclination angle $i\neq0^\circ$: In the lower edge of the outflow (left and middle panels), the molecular outflow 
has velocities of the order to 12 km s$^{-1}$, this high velocity respect to the local standard of the rest velocity $V_{\rm LSR}$ = 5 km s$^{-1}$ (\citealt{Plambeck_2016}), can be explained as the axial velocity. Here, we assume an inclination for the outflow of $i=$10$^\circ$, which is similar to the value reported by \citet{Plambeck_2016}, \citet{Hirota_2017}, and \citet{Baez_2018}.
In addition, in the panels (a) and (b), one can observe that the size of the molecular outflow is $\sim$ 1400 au. 
Finally, the panel (c) show that the molecular line of $^{28}$SiO (J=8--7) $\nu$=1 traces the inner most part of the molecular outflow of Orion Src I.
Figure \ref{fig:mom1-29sio} also shows the 1.3 mm continuum emission (in white contours) from Orion Src I. This continuum emission is tracing the disk surrounding this source, see 
\citet{Hirota_2017,Plambeck_2016}. 

The position-velocity diagrams of the emission from the molecular line of $^{29}$SiO (J=8--7) $\nu$=0 are shown in Figure \ref{fig:pv29sio}. 
This Figure presents parallel cuts at different distances from the disk mid-plane, these cuts were made from $z$=480 au to $z$=$-$480 au 
with intervals of 80 au (see the dashed lines in panel (a) of Figure \ref{fig:mom1-29sio}). One can observe that in regions near to the disk, this molecule fills the molecular outflow, while, for regions far from the disk, this molecule presents a thin--shell structure
in expansion. In addition, one can observe that all position-velocity diagrams present signatures of the rotation (see panel a of Figure \ref{fig:pvrot}).

In Figure \ref{fig:pvsis} we have done a similar analysis to Figure \ref{fig:pv29sio} for the emission from the molecular line of SiS (J=19--18) $\nu$=0. 
The position-velocity diagrams show a thin shell structure where the emission from this molecule is very prominent. The width of the shell is $\Delta r \sim$ 120 au which is $\sim$ 1/3 of the distance to the central star. This molecule shows that
the outflow is in expansion because the size of the thin shells increases with the distance from the disk. In these diagrams the rotation 
of the molecular outflow is confirmed. The biggest rotation velocity corresponds
to a height of $z=\pm$80 au (see panel b of Figure \ref{fig:pvrot}).

Figure \ref{fig:pvsio} shows the position-velocity diagrams of the emission from the molecular line $^{28}$SiO (J=8--7) $\nu$=1 for the same 
distances from the disk mid-plane of the Figures \ref{fig:pv29sio} and \ref{fig:pvsis}. In contrast to the other two molecules, 
in this molecular line the thin shell structure does not appear. This Figure confirms the presence of the rotation in the molecular outflow (see panel c of Figure \ref{fig:pvrot}). 
Finally, the absence of the emission for distances of $z\ge \pm$320 au means that this molecule is only tracing the inner part of the molecular outflow. This is maybe due to excitation conditions. 

\citet{Hirota_2017} measured the rotation velocities for heights between $z=-200$ au and $z=200$ au, and they found that these velocities decrease with the height and have values between $\sim$3--9 km s$^{-1}$. In this work, we reported rotation velocities for the same heights of the order of 4--8 km s$^{-1}$, these values are similar to those reported by these authors.

Finally, Figure \ref{fig:pvrot} clearly shows the evidence of the rotation and the expansion in Orion Src I. \footnote{If the gas is expanding and rotating, the position velocity diagrams show an elliptical structure with the semi major axis inclined with respect to the position axis 
(see, e.g, panel d of the Supplementary Figure 1 of \citealt{Hirota_2017}).}. 
In this figure, we have made a zoom to the position-velocity 
diagrams of the Figures \ref{fig:pv29sio}, \ref{fig:pvsis}, and \ref{fig:pvsio} at a distance of $z=-$80 au from the disk for the molecular 
lines of $^{29}$SiO (J=8--7) $\nu$=0 (panel a), SiS (J=19--18) $\nu$=0 (panel b), and $^{28}$SiO (J=8--7) $\nu$=1 (panel c), respectively.

\begin{figure*}[!t]
\centering
\includegraphics[scale=0.55]{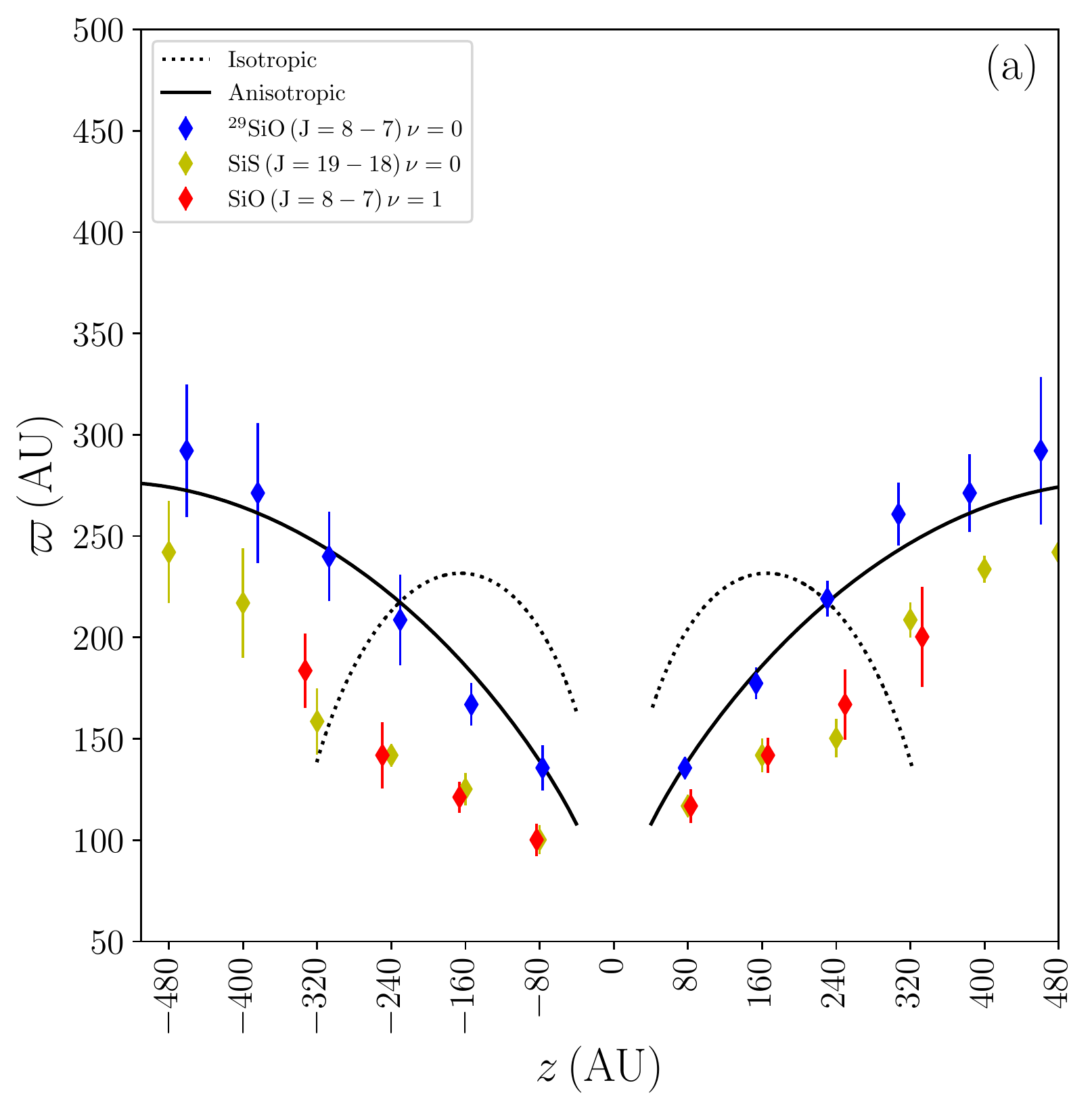}\includegraphics[scale=0.55]{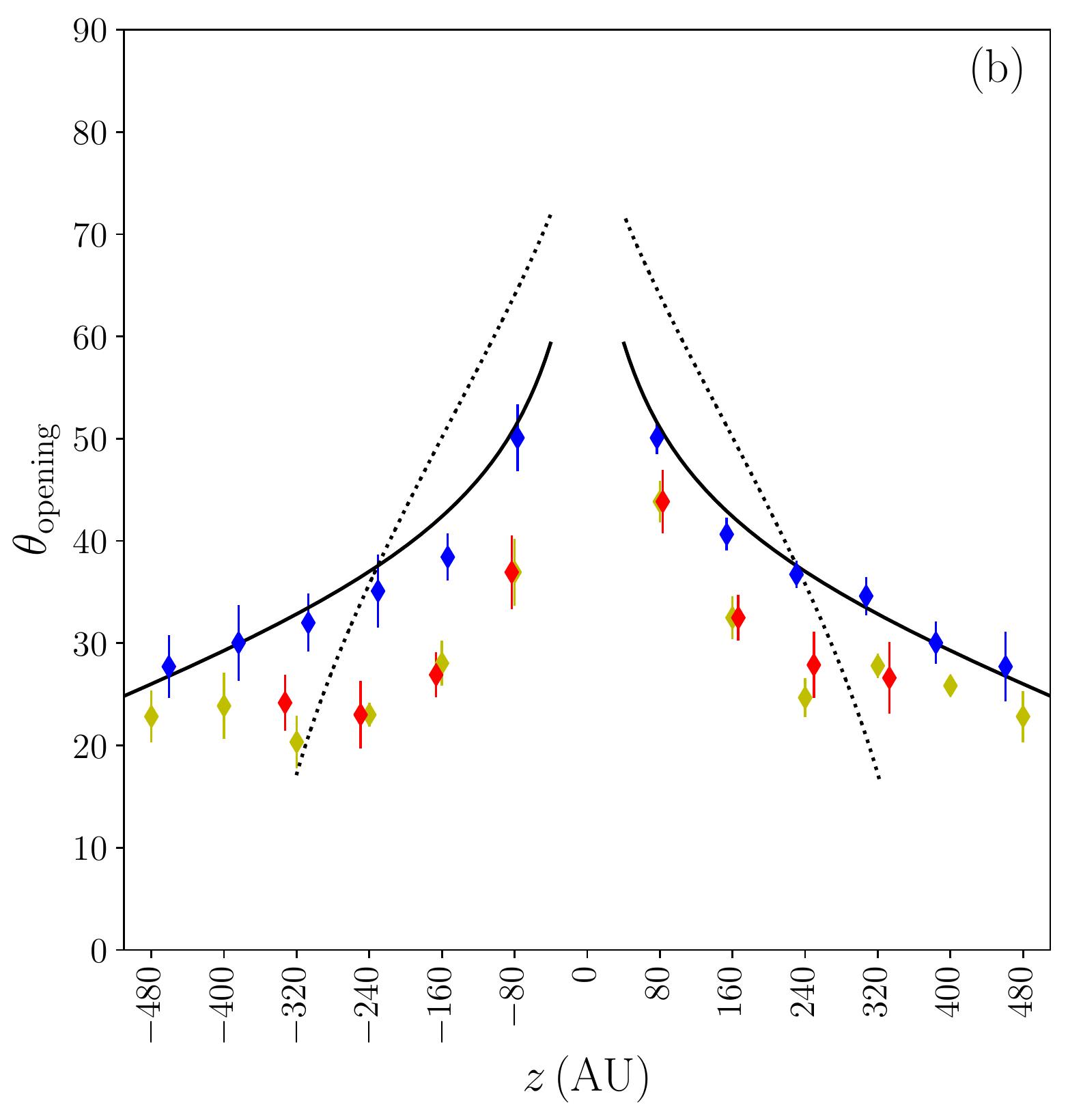}
\caption{
Panel (a) shows the  cylindrical radii of the outflow $\varpi$; panel (b) shows the opening angle of the outflow $\theta_{\rm opening}$. 
These observed values are derived from the position-velocity diagrams  in Figures \ref{fig:pv29sio}, \ref{fig:pvsis}, and \ref{fig:pvsio}. 
The error bars are derived from the gaussian fit (see Appendix \ref{app:complementary} for the measurement procedure).
The dotted line shows an isotropic model with the parameters $\alpha=0.1$, $\beta=4$, $r_{s0}(0)=10^{-3}$, and $B=0$.
The black line shows the best anisotropic stellar wind model with the parameters $\alpha=0.1$, $\beta=4$, $r_{s0}(0)=10^{-3}$, $A=1$, $B=35$, and $n=5$. Both models were integrated up to a dynamical time of 65 yr.}
\label{fig:graphicsr}
\end{figure*}

\begin{figure*}[!t]
\centering
\includegraphics[scale=0.375]{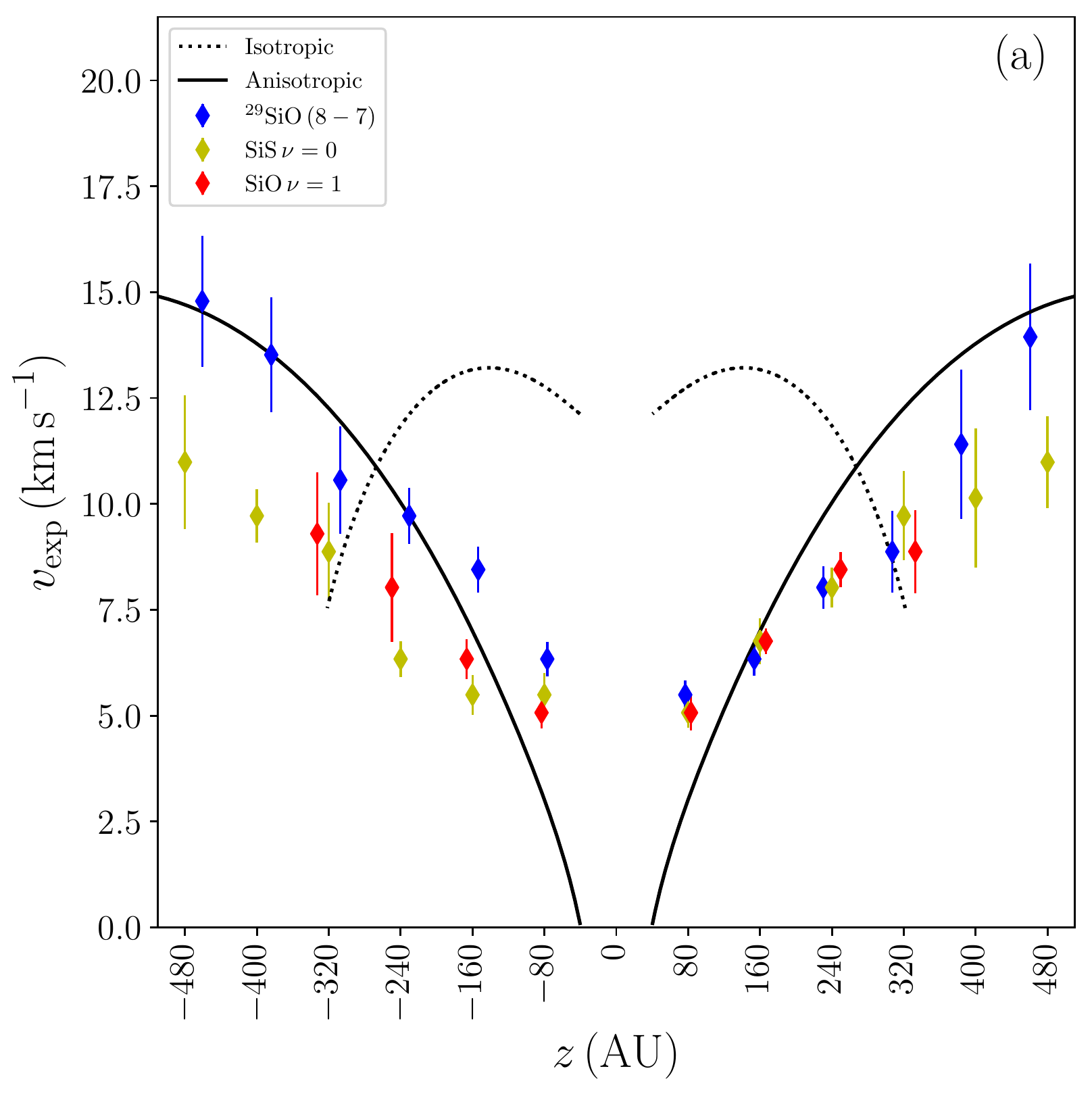}\includegraphics[scale=0.375]{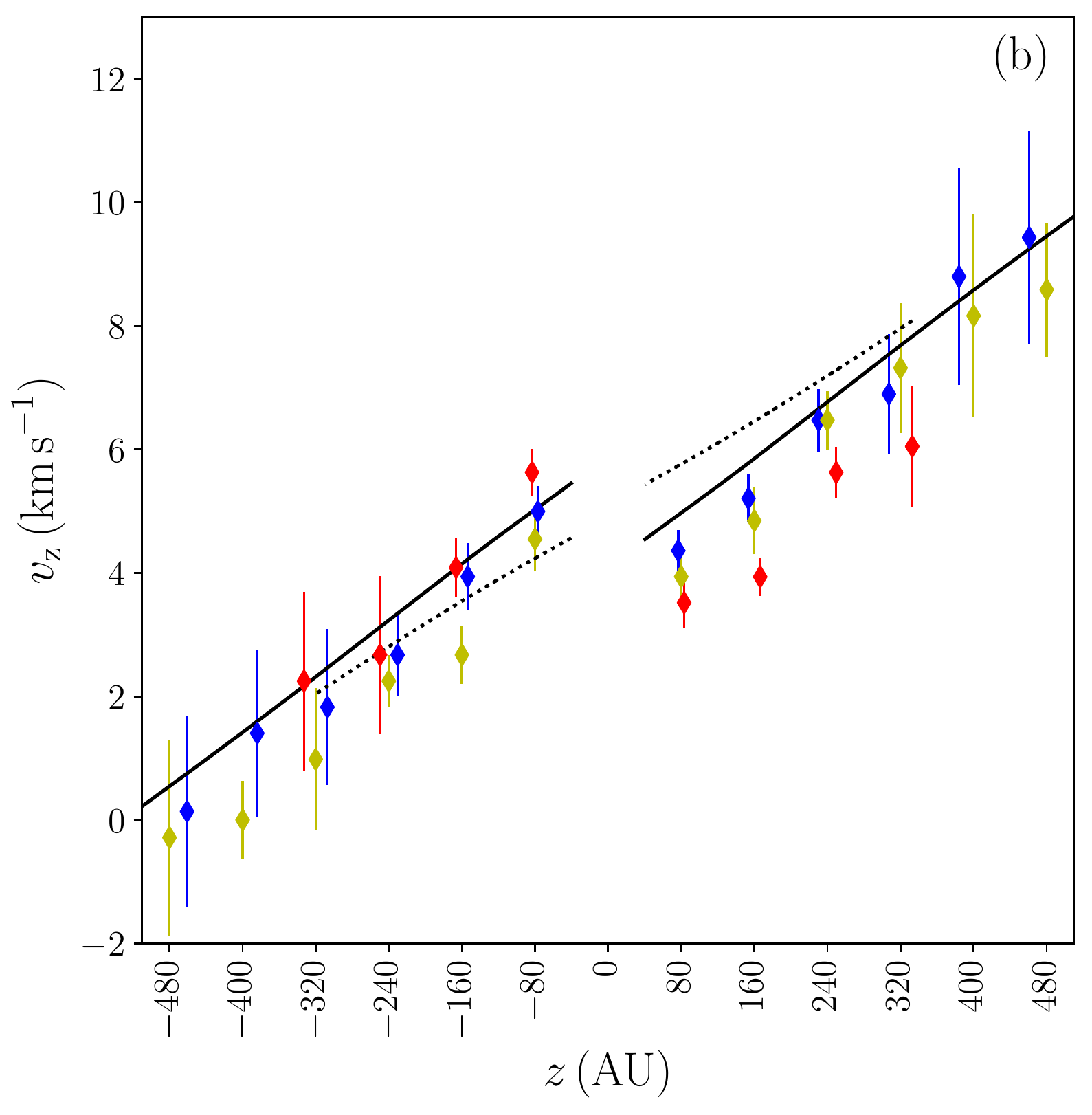}\includegraphics[scale=0.375]{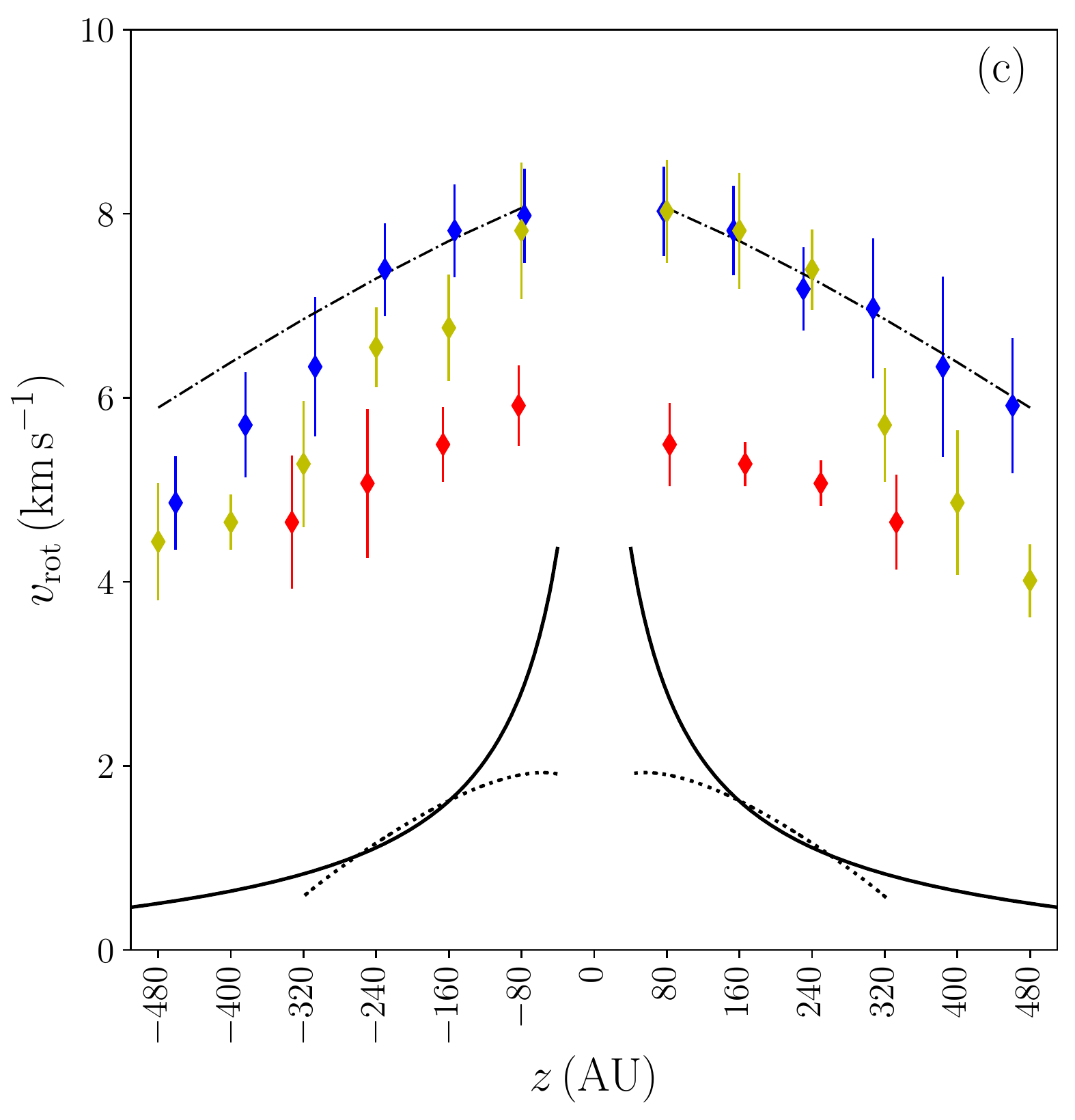}
\caption{Panel (a) shows the expansion velocity perpendicular to the outflow axis $v_{\rm exp}$ measured at the cylindrical radius;
panel (b) shows the axial velocity $v_{\rm z}$; panel (c) shows the rotation velocity $v_{\rm rot}$ measured at the cylindrical radii.
These observed values are derived from the position-velocity diagrams  in Figures \ref{fig:pv29sio}, \ref{fig:pvsis}, and \ref{fig:pvsio}. 
The error bars are derived from the gaussian fit (see Appendix \ref{app:complementary} for the measurement procedure).
The dotted line shows the isotropic model with the parameters of Figure \ref{fig:graphicsr}.
The black line shows the best anisotropic stellar wind model with the parameters of Figure \ref{fig:graphicsr}.
The dashed dotted line of the panel (c) corresponds to the best fitting of the function $v_{\rm rot}=a z^{\gamma}+b$ (see text).}
\label{fig:graphicsv}
\end{figure*}

\subsection{Mass of the outflow}
\label{subsec:mass}
Assuming that the $^{28}$SiO (J=8-7) $\nu=1$ emission is optically thick, the excitation temperature is (e.g., 
\citealt{Estalella_1994})
\begin{eqnarray}
T_{\rm ex}(^{28}{\rm SiO})=\frac{h\nu/k}{\mathrm{ln}\left(1+\frac{h\nu/k}{T_{\rm a}(^{28}{\rm SiO})+J_\nu(T_{\rm bg})}\right)},
\label{eq:tex}
 \end{eqnarray}
where $h$ is the Plank constant, $k$ is the Boltzmann constant, $\nu$ is the rest frequency in GHz (see Table \ref{tab:molecules}), $T_{\rm a}(^{28}{\rm SiO}) = 19$ K is the observed antenna temperature of $^{28}$SiO, and  $J_\nu(T_{\rm bg})$ is intensity in units of temperature at the background temperature $T_{bg}=2.7$ K. Using  the value of $\nu$ given in Table \ref{tab:molecules}, we obtain $T_{\rm ex}(^{28}{\rm SiO})  =  26$ K. Assuming that the $^{28}$SiO and $^{29}$SiO molecules coexist
and share the same excitation temperature, $T_{\rm ex}(^{28}{\rm SiO})=T_{\rm ex}(^{29}{\rm SiO})=T_{\rm ex}$, we can estimate the optical depth of the $^{29}$SiO molecule as (e.g., 
\citealt{Estalella_1994})

\begin{eqnarray}
\tau_0(^{29}{\rm SiO})=-{\rm ln}\left[1-\frac{T_{\rm a}(^{29}{\rm SiO})}{J_\nu(T_{\rm ex})-J_\nu(T_{\rm bg})}\right],
\label{eq:opticaldepth}
\end{eqnarray}
where $T_{\rm a}(^{29}{\rm SiO})=14$ K is the observed antenna temperature of $^{29}$SiO and $J_\nu(T_{\rm ex})$ is the intensity in units of temperature at the excitation temperature. With these values, we obtain $\tau_0(^{29}{\rm SiO})=1.3$, which is not optically thin. 
Thus, assuming local thermodynamic equilibrium, we calculate the mass of the outflow as a function of 
the $^{29}$SiO optical depth as
\begin{eqnarray}
\frac{\mathrm{M}_{\rm outflow}}{\mathrm{M}_\odot}&=&5\times10^{-23}\left(d^2 \Delta\Omega \right)\left[\frac{m(\mathrm{H_2})}{X\left ( \frac{ ^{29}{\rm SiO}  } {{\rm H}_2} \right )} \right] \nonumber \\
&\times&\frac{\mathrm{exp}\left[\frac{58.6}{T_{\rm ex}}\right]}{1-\mathrm{exp}\left[\frac{-16.7}{T_{\rm ex}}\right]} 
T_{\rm ex} \tau_0 (^{29}{\rm SiO}) \Delta v,
\label{eq:massout}
\end{eqnarray}
where $m$(H$_2$) is the mass of the molecular hydrogen, 
 $\mathrm{X}\left ( \frac{ ^{29}{\rm SiO}}{{\rm H}_2} \right )= 6 -12 \times 10^{-9} $ is the fractional abundance of $^{29}$SiO with respect to
 H$_2$\footnote{ The factors 58.6 and 16.7 in this equation, are the result of $4.8\times10^{-2}\times B_{e}J(J+1)$ and $4.8\times10^{-2}\times 2B_{e}(J+1)$, respectively, where $B_{e}=21.8$ GHz is the rotational constant of the molecule $^{29}$SiO, $J=7$ is the lower level, the factor of $4.8\times10^{-2}$ is the ratio of $h/k$ in GHz$^{-1}$.}.
 To obtained this value, we assumed a relative abundance of $^{28}$SiO with respect to H$_2$ of $1.2 -2.4  \times 10^{-7}$,
 obtained by \citet{Ziurys_1987} in OMC1 (IRc2), and  a relative abundance of $^{29}$SiO with respect to $^{28}$SiO of
  5$\times$10$^{-2}$, obtained by \citet{Soria_2005} toward evolved stars.  The distance $d$ is  (418$\pm$6 pc), $\Delta v$ is the velocity width of the line  ($\sim$30 km s$^{-1}$), and $\Delta \Omega$ is the solid angle of the source ($\sim1.33\times 10^{-9}$ sr). With these values, the estimated  mass of the outflow of Orion Src I is $\rm{M}_{\rm outflow}\gtrsim 0.66 - 1.3$ M$_\odot$.
This mass  is a lower limit because 
 the $^{28}$SiO abundance could be lower by up to two orders of magnitude due to the uncertainty in the molecular hydrogen column densities (\citealt{Ziurys_1987}).

In addition, for an expansion velocity $ v \sim 18\  {\rm km \ s^{-1}}$ \citep{Greenhill_2013} and a size  $z= 480$ au,  
the kinematic time  is $t_{kin} \sim 130 $ yr. 
Then, the mass loss rate of the molecular outflow as $\dot{\rm M}_\mathrm{outflow}=
\rm M_{\rm outflow}/t_{kin}\gtrsim 5.1 - 10  \times10^{-3}$ M$_\odot$ yr$^{-1}$.

 \citet{Hirota_2017} proposed that molecular outflow of Orion Src I 
 is produced by a slow magnetocentrifugal disk wind. The observed values of the rotational velocities of the
 outflow can be reproduced by this model which predicts that the wind is eject from footpoints in the disk at radii $ r \sim 5 -  25$ au.

A disk wind requires a very large mass loss rate to account for the mass observed in the outflow. 
  As mentioned in the Introduction, recent MHD simulations show that disk winds around T Tauri stars
  can have $\dot M_{\rm w} = f \dot M_{\rm d,a}$, where the fraction can be 
  $f \sim 1 -2 $ (e.g., \citealt{Bai_2017}; \citealt{Bethune_2017}; \citealt{Wang_2019}). 
  If the outflow is a disk wind,  $\dot M_{\rm outflow} = \dot M_{\rm w}$. 
  In the case of Orion SrcI, this implies a very large disk accretion rate,
  $f \dot M_{\rm d,a}  \gtrsim (5.1- 10)  \times 10^{-3} M_\odot {\rm yr}^{-1}$.  
Then,  massive disk winds face two problems.
The first problem has to do with the
 fact that the mass flux in the disk will eventually fall into the star.  Assuming that the disk rotates with 
 Keplerian speed $v_K$, the material accreted to the star has to dissipate its energy, $1/2 \dot M_{\rm d,a} v_K^2$. Thus, 
    the accretion luminosity at the stellar surface is given by  
 $L_{\rm a} = \eta {G M_* \dot M_{d,a} \over R_*}$, where $G$ is the gravitational constant, $M_*$ is the stellar mass, $R_*$ is the stellar radius, 
 and $\eta \sim 0.5$. Assuming $M_* = 15 M_\odot$ (\citealt{Ginsburg_2018}) and $R_* = 7.4\,R_\odot$ (\citealt{Testi_2010}), the accretion luminosity is
 $L_{\rm a}  \gtrsim  (1/f) 1.5 \times 10^5 L_\odot$. This value is higher than 
 the observed source luminosity $L_* \sim  10^4 L_\odot$ (e.g., \citealt{Menten_1995}; \citealt{Reid_2007}),  unless $f \sim 15$.
Note that a  factor $f \sim 15 $ implies that (locally)   94\% of the mass 
 the mass escapes into the wind and only  6\% accretes towards the star. 
 Disk wind models would have to produce these high $f$ values in the case of winds around massive stars. 
 The second problem, that was already mentioned in the case of DG Tau B (Section \ref{sec:introduction}), is the short disk lifetime.
 For a maximum disk mass $M_d \lesssim M_*/3 = 5 M_\odot$, necessary for gravitational stability (\citealt{Shu_1991}),  and 
 an accretion rate such that $f \dot  M_{\rm d,a}  \gtrsim 5.1 - 10 \times 10^{-3} M_\odot {\rm yr}^{-1}$, the disk lifetime is very small, 
 $\tau = M_d / \dot M_{\rm d, a} \lesssim  f \times 980 \, {\rm yr}$ (see also the short disk lifetimes in Fig. 33 of \citealt{Bethune_2017} for disks around low mass stars).  This estimate of the disk lifetime assumes that the disk mass is not replenished. Nevertheless, Orion Src I has a massive accreting envelope that could replenish the disk. 
The disk wind models would have to explore if the disk mass could be
 replenished in short timescales ($ \lesssim  10^4$ yr)  by the infalling envelope. Both,  the accretion luminosity and the disk lifetime, 
 are important constraints on the disk wind models. 
 
Moreover, if there is an accreting envelope around the Orion Src I, 
a stellar or disk wind will necessarily collide against it, driving a shell of entrained material. 
For this reason, in this work we explore  a model where  the molecular outflow is a shell produced by the interaction of a stellar wind and an accretion flow as the scenario first proposed by \citet{Snell_1980}.
The shell is fed by both the stellar wind and the accretion flow. The latter can
 have very large mass accretion rates as observed in the case of young massive stars  
 (e.g., \citealt{Zapata_2008}; \citealt{Wu_2009}). We will verify under which conditions this shell model
 can acquire the observed mass.

\subsection{Comparison with the outflow model}
\label{subsec:model}

The position-velocity diagrams, presented in Figures \ref{fig:pv29sio}-\ref{fig:pvsio}, show the detailed structure of the outflow velocity as 
a function of the distance from the disk mid--plane. With these diagrams, we can also obtain information about the kinematic 
and physical properties of the outflow and compare with the outflow model of LV19.

\citet{Goddi_2011} suggested that this source is a binary system with a stellar mass of $\sim$20 M$_\odot$, 
and a separation of the stars $<$10 au. Since this separation is very small compared to the size of the outflow, 
even if each star has its own stellar wind, a single stellar wind emanating from the center is a good approximation.

The proper motion of the Orion Src I with respect to the center of the explosive event that occured 500 yr ago (\citealt{Rodriguez_2017})
will change the environment of the central star. Its envelope will not be a gravitational collapsing envelope of the Ulrich type 
since the free fall time of a gas parcel starting at an outflow distance $\sim 1000$ au from star is of the
order of twice the crossing time. Nevertheless, we will apply the models of LV19 and see how well the observational properties of the
outflow can be reproduced.

The model of LV19 assumes that the molecular outflow is a thin shell formed by the collision between a stellar 
wind and a molecular rotating cloud in gravitational collapse. 
The thin shell assumption is adequate because the width of the shell is $\Delta r \sim 1/3 $ of the distance to central star (see Section \ref{subsec:observations}).
For our comparison we assume a stellar mass M$_*=15$ M$_\odot$ (\citealt{Ginsburg_2018}) and
a centrifugal radius of $R_{\rm cen}=40$ au, within the range of 21 au  - 47 au reported 
by \citealt{Hirota_2017}. 

This model depends of two parameters associated with the properties of the stellar wind and the accretion flow. 
The first parameter is the ratio between the wind mass loss rate $\dot{\mathrm{M}}_w$, and the mass accretion rate 
of the envelope $\dot{\mathrm{M}}_a$

\begin{eqnarray}
\alpha=\frac{\dot{\mathrm{ M}}_w}{\dot{\mathrm{M}}_{a}},
\label{eq:alpha}
\end{eqnarray}

\noindent for this case, we assume a value of $\alpha=0.1$, a typical value the molecular outflows (\citealt{Ellerbroek_2013}; \citealt{Nisini_2018}). 
The second parameter is the ratio between the stellar wind and the accretion flow momentum rates

\begin{eqnarray}
\beta=\frac{\dot{\mathrm{M}}_w v_w}{\dot{\mathrm{M}}_{a} v_0} = \alpha \frac{v_w}{v_0},
\label{eq:beta}
\end{eqnarray} 
where $v_w$ is the velocity of the stellar wind, and $v_0$ is the free fall velocity at the centrifugal radius, given by
\begin{eqnarray}
v_0=\left(\frac{\mathrm{GM}_*}{R_{\rm cen}}\right)^{1/2}.
\label{eq:v0}
\end{eqnarray}
For inferred values $\mathrm{M}_* = 15\,\mathrm{M}_\odot$ and $R_{\rm cen} = 40$ au, 
 the free fall velocity is $v_0=19$ km s$^{-1}$. Assuming a stellar 
wind velocity  $\sim 800$ km s$^{-1}$, of the order of the escape speed for a star with $R_* \sim 7.4 R_\odot$ (\citealt{Testi_2010}),
implies that  $\beta\simeq4$. 

We assume a density profile of the stellar wind given by 
\begin{eqnarray}
\rho_w=\frac{\dot{\mathrm{M}}_w}{4\pi r^2v_w}f\left(\theta\right),
\label{eq:rhow}
\end{eqnarray}
\noindent where $f\left(\theta\right)$ is the anisotropy function given by
\begin{eqnarray}
f\left(\theta\right)=\frac{A+B\cos^{2n}\theta}{A+B/(2n+1)}.
\label{eq:fanisotropy}
\end{eqnarray}

The physical properties of the shell model  that will be compared with the observations are: 
 the cylindrical radius $\varpi$, the opening angle $\theta_{\rm opening}$, the expansion velocity $v_{\rm exp}$, the axial velocity $v_z$, and the rotation velocity $v_{\rm rot}$. Figure \ref{fig:esquema} presents a schematic 
diagram of the molecular outflow that shows the cylindrical radius, the height over the disk mid-plane, 
and the opening angle. The procedure used to measured these quantities is described in Appendix \ref{app:complementary}.

We considered two models:  a shell formed by 
an isotropic stellar wind with $B=0$;
and a shell formed by a very anisotropic stellar wind, with 
 $A=1$, $B=35$, and $n=5$. 
 The parameters of the anisotropic stellar wind model are chosen to reproduce the 
shape of  the most extended outflow emission as traced by the $^{29}$SiO (J=8--7) $\nu=0$ transition.
We choose the parameters that minimize $\chi^2$, defined as
\begin{eqnarray}
\chi^2=\frac{1}{N}\sum\frac{\left(\varpi_{\rm obs}-\varpi_{\rm model}\right)^2}{\varpi_{\rm obs}^2},
\label{eq:chi}
\end{eqnarray}
where $\varpi_{\rm obs}$ is the observed cylindrical radius, $\varpi_{\rm model}$ is the model cylindrical radius, and $N$ is the number of observed values along the $z$ axis. 
This analysis is shown in Figure \ref{fig:chisq}.
We integrate in time the shell model from a small initial shell radius  $r_{s0}(0)\simeq R_*/R_{\rm cen}\sim 10^{-3}$, close to the stellar surface,
until the cylindrical radii of the model $\varpi_{\rm model}$ 
reaches the observed cylindrical radii $\varpi_{\rm obs}$ at different heights as shown in 
panel (a) of the Figure \ref{fig:graphicsr}, which happens at $t = 65$ yr. The shell model $R_s(\theta)$ is
shown in Figure \ref{fig:bestmodel}. 
 Because the shell decelerates with time, the dynamical time (65 yr) is half of the kinematic time (130 yr)
calculated in Section \ref{subsec:mass}.
Figure \ref{fig:bestmodel} shows shell produced by the isotropic wind (dashed line) and the anisotropic stellar wind (solid line) 
model superimposed on the ALMA first moment of the line emission $^{29}$SiO (J=8--7) $\nu$=0 (panel a), 
SiS (J=19--18) $\nu$=0 (panel b), and $^{28}$SiO (J=8--7) $\nu$=1 (panel c).

The comparison between both outflow models with the observational data is shown in 
 Figures \ref{fig:graphicsr} and \ref{fig:graphicsv}. Since the isotropic wind model (dotted lines) does not reproduce 
the observations, hereafter, we will only discuss the properties of the anisotropic stellar wind model. 

The panel (a) of the Figure \ref{fig:graphicsr} shows the cylindrical radius obtained from the three line 
observations and from the anisotropic stellar wind model. These radii are shown as vertical dashed lines in Figure \ref{fig:pvrot}. 
The cylindrical radius $\varpi$ increases with the height above the 
disk mid-plane, and one can see that the model (black solid lines) agree well with observational data.

For fixed centrifugal radius $R_{\rm cen}$, the opening angle can be defined as
\begin{eqnarray}
\theta_{\rm opening}=\tan^{-1}\left(\frac{\varpi-R_{\rm cen}}{z}\right).
\label{eq:thetaop}
\end{eqnarray}
This angle is shown in panel (b) of the Figure \ref{fig:graphicsr}.  The observed values and  the model
(black solid lines)  are consistent.
The fact that the opening angle decreases with the height above the disk, 
 indicates that the molecular outflow could close up at higher heights. Nevertheless, one needs observations of a molecule
that emits at higher disk heights to establish the outflow shape.

The panel (a) of the Figure \ref{fig:graphicsv} shows the expansion velocity for the three molecules indicated in the panel. 
This velocity increases 
with the height above the disk mid-plane. The model expansion velocities (black solid lines) 
are similar to the observed values 
except close to the disk ($z <  \pm 150 $ au). 

Panel (b) of the Figure \ref{fig:graphicsv} shows the  measured axial velocity $v_z$. 
This velocity increases with the height above the disk mid--plane. The axial velocity of the 
anisotropic stellar wind model corrected by the inclination angle $i = 10^\circ$ and a system velocity
$V_{\rm LSR} = 5 \ {\rm km\,s}^{-1}$ (e.g., \citealt{Plambeck_2016}) fits the data well.

The rotation velocity is shown in panel (c) of the Figure \ref{fig:graphicsv}. For the molecular line of $^{29}$SiO (J=8--7) $\nu=0$ (blue points), the rotation velocity is in the range  5--8 km s$^{-1}$: at
$z=\pm$80 au above the disk the rotation velocity is $\sim$8 km s$^{-1}$ and it decreases with height. The SiS (J=19--18) $\nu=0$ line (yellow points) has a similar behavior. The $^{28}$SiO (J=8--7) $\nu=1$ emission
 (red points) behaves in the same way but has slightly lower velocities, in the range 4--6 km s$^{-1}$.
  The observed rotation velocity is a factor
of 3$-$10 larger than those of the anisotropic stellar wind model. Furthermore, the rotation velocity of the model decreases 
steeply with the height; the observed rotation velocity slowly decreases. For  reference,
a polynomial function $(v_{\rm rot}/ \rm{km s}^{-1})=a \left(z / {\rm au}\right)^{\gamma}+b$ with $a=-1.5\times10^{-3}$, $\gamma=1.2$, and $b=8.3$  
is shown  as a dashed dotted line in panel (c) of Figure \ref{fig:graphicsv}.

One can also compare the model shell mass  with the observed outflow mass (Section \ref{subsec:mass}). The shell mass is given by  
\begin{eqnarray}
M_{\rm outflow}= \frac{\dot M_{\rm a} R_{\rm cen}}{v_0} \int_0^{\pi/2} p_m d\theta,
\label{eq:mass}
\end{eqnarray}
where $p_m$ is the non dimensional mass flux (eq. [47] of LV19). For the anisotropic stellar wind model, $\int_0^{\pi/2} p_m d\theta =6.0$, in non
dimensional units. Therefore, for the
values of the centrifugal radius and free fall velocity above, one requires 
 a mass accretion rate of the envelope $\dot{\mathrm{M}}_a = 1.1 - 2.2 \times10^{-2}$ M$_\odot$ yr$^{-1}$  to obtain the observed mass of the shell, $M_{\rm outflow} = 0.66 - 1.3 M_\odot$.
  Such large mass envelope accretion rates have been inferred in regions of high mass star formation (e.g., \citealt{Zapata_2008}; \citealt{Wu_2009}.)
 This accretion rate corresponds a mass loss rate of the molecular outflow corrected by the dynamical time $\dot M_{\rm outflow}^\prime = (0.66 - 1.3$ M$_\odot) / 65$ yr
 = 1 - 2 $\times 10^{-2}$ M$_\odot$ yr$^{-1}$ which is very similar to 
 $\dot{M}_{\rm a} $. Thus, the small fraction of mass that slides along the shell towards the 
 equator does not increase the disk mass significantly.

In summary, the comparison between the anisotropic stellar wind model and the observations of  the outflow from Orion Src I fits very
well the outflow cylindrical radius. The opening angle is a function of the cylindrical radius, therefore, it also fits the observations well. The expansion velocity and the axial velocity $v_z$ have a behavior similar to the observations, although the slope is somewhat different. Nevertheless, the model rotation velocity is much lower (3$-$10 times) than the observed velocity.
 
The smaller rotation velocity profile of the model indicates that the envelope of \citet{Ulrich_1976} 
can not explain the rotation in molecular outflows. This problem could be alleviated if
 one includes a stellar wind or disk wind with angular momentum, or increases
the angular momentum of the envelope. 

 For a representative height of $z  \sim 240 $ au, the observed rotation velocity is a factor 
$\sim 6$ of the model rotation velocity. Thus, the  model has only $\sim 17 \%$ of the 
observed specific angular momentum.
 The missing angular momentum could come from an accreting envelope with 
 more angular momentum, or from an extended disk wind. \footnote{An X wind comes from radii very close to the central star, so it has little angular momentum.}
 
\section{Conclusions}
\label{sec:Conclusions}

In this study, we present new and sensitive ALMA archive observations of the rotating outflow from Orion Src I.
 In the following, we describe our 
main results.  

\begin{itemize}

\item The Orion Src I outflow has a mass loss rate $\dot M_{\rm outflow} = 5.1 - 10\times10^{-3} \, M_\odot {\rm yr}^{-1}$. This massive outflow poses stringent constraints on disk wind models 
concerning  the accretion luminosity and the disk lifetime. 

\item We find that the opening angle (in a range of $\sim$20--60$^\circ$) and
the rotation velocity (in a range of $\sim$4--8 km s$^{-1}$) decrease with the height to the disk. In contrast,
the cylindrical radius (in a range of $\sim$100--300 au), the expansion velocity (in a range of $\sim$2--15 km s$^{-1}$), and the axial velocity $v_{\rm z}$ (in a range of $\sim$ -1--10 km s$^{-1}$)
increase with respect to the height above the disk.

\item We compare with the outflow model of LV19,  where the molecular outflow corresponds to a shell produced by the interaction of a stellar
wind and an accretion flow.

\item We find that the observed values of the cylindrical radius, the opening angle, the expansion velocity, and the axial velocity $v_{\rm z}$ show a similar behavior to LV19 anisotropic stellar wind model. 
However, the rotation velocity of the model is lower (by a factor of 3--10) than the observed rotation velocity 
of the Orion Src I outflow.

\item We conclude that the Ulrich flow alone cannot explain the rotation of the molecular
         outflow originated from Orion Src I and other possibilities should be explored. 

\end{itemize}

We thank the referee for very useful comments that improved the presentation of the paper.
J.A. L\'opez-V\'azquez and Susana Lizano acknowledge support from PAPIIT--UNAM IN101418 and CONACyT 23863. 
Luis A. Zapata acknowledges financial support from DGAPA, UNAM, and CONACyT, M\'exico. Jorge Cant\'o acknowledges support from PAPIIT--UNAM--IG 100218.
This paper makes use of the following ALMA data:  ADS/JAO.ALMA\#2016.1.00165.S. and  ADS/JAO.ALMA\#2012.1.00123.S
ALMA is a partnership of ESO (representing its member states), NSF (USA) and NINS (Japan), 
together with NRC (Canada), MOST and ASIAA (Taiwan), and KASI (Republic of Korea), 
in cooperation with the Republic of Chile. The Joint ALMA Observatory is operated 
by ESO, AUI/NRAO and NAOJ. In addition, publications from NA authors must include 
the standard NRAO acknowledgement: The National Radio Astronomy Observatory is a 
facility of the National Science Foundation operated under cooperative agreement 
by Associated Universities, Inc.

\appendix
\section{Measurement procedure}
\label{app:complementary}

\begin{figure*}[t!]
\centering
\includegraphics[scale=0.34]{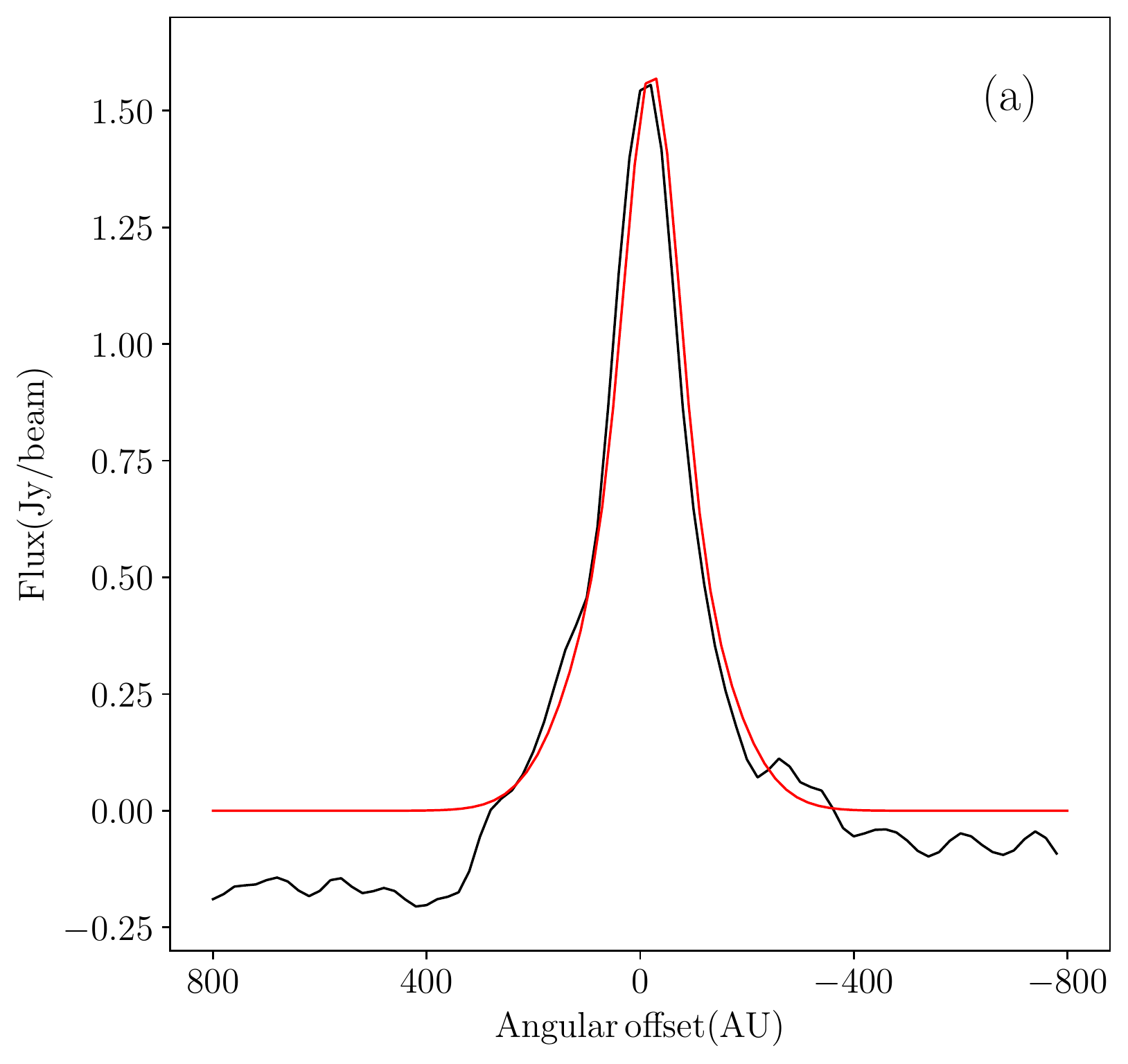}\includegraphics[scale=0.34]{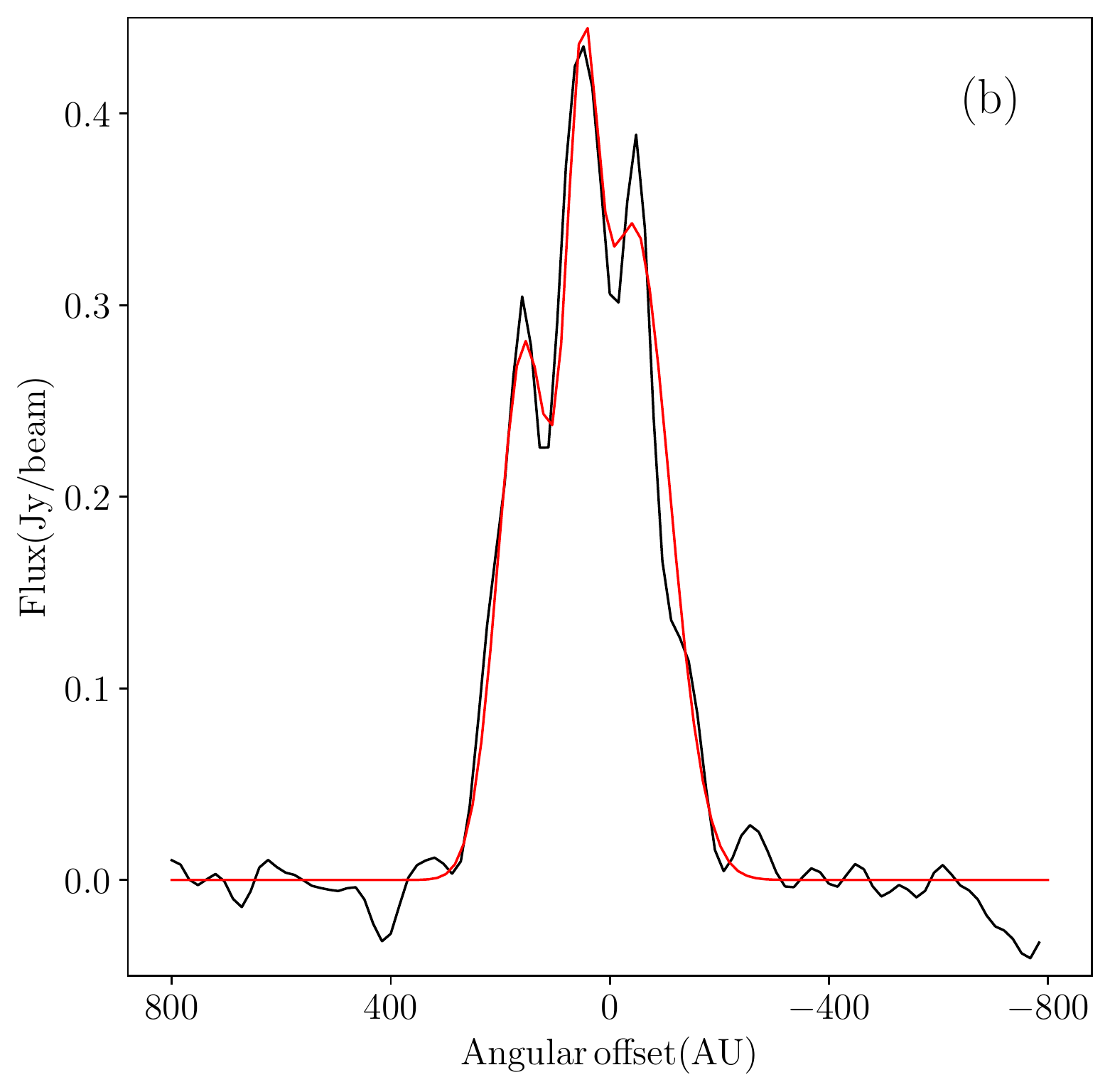}\includegraphics[scale=0.34]{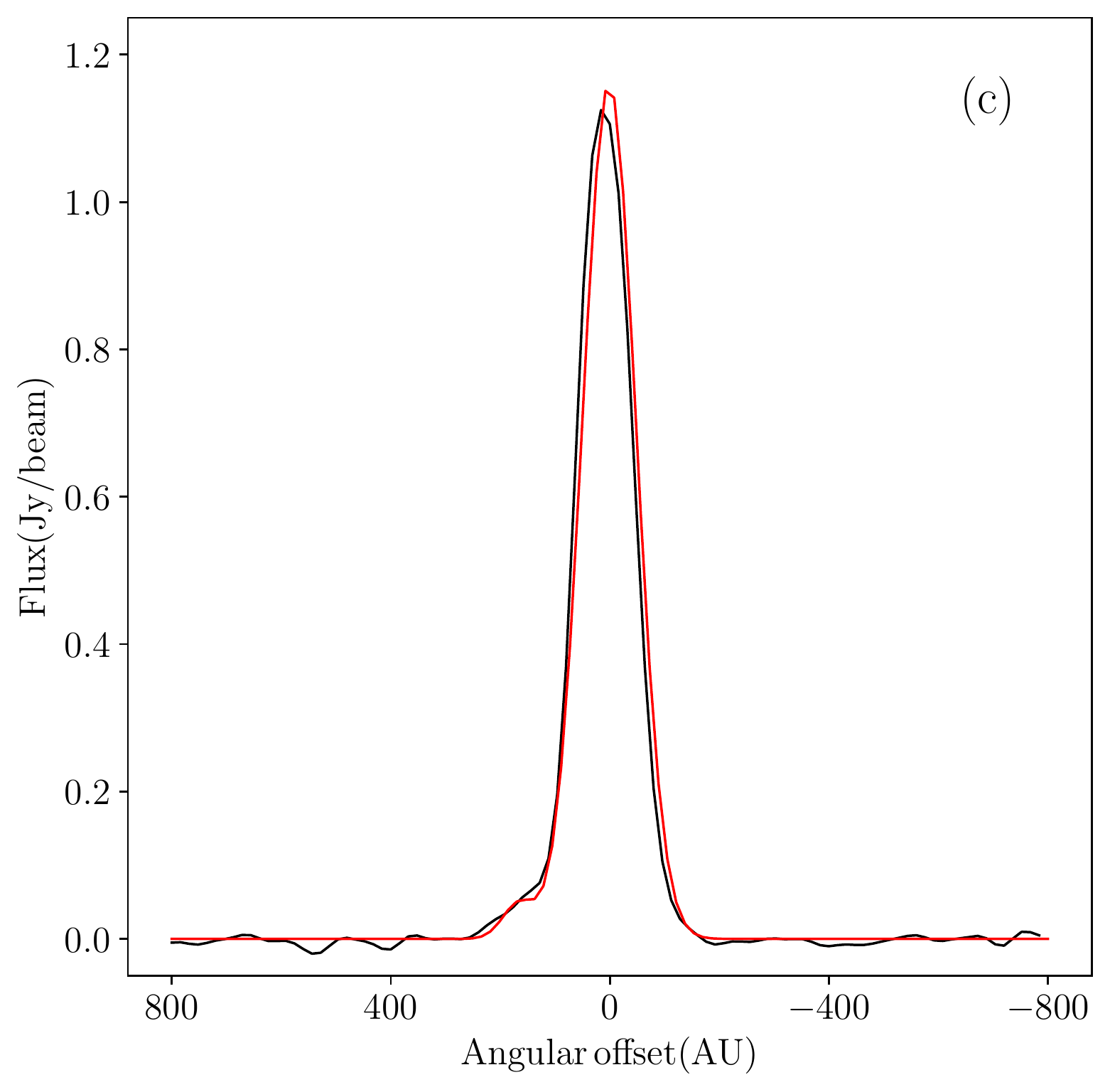}
\caption{Intensity profiles  at $V_{\rm LSR}=5$ km s$^{-1}$ of the position-velocity diagrams at a height $z=-80$ au,
 as indicated by horizontal dashed lines in panels (a)--(c) in Figure \ref{fig:pvrot}. The red line shows the best Gaussian fit to the intensity profile of
 (a) $^{29}$SiO (J=8--7) $\nu=0$, (b) SiS (J=19--18) $\nu=0$, and
(c) $^{28}$SiO (J=8--7) $\nu=1$. }
\label{fig:routapp}
\end{figure*}

\begin{figure*}[t!]
\centering
\includegraphics[scale=0.34]{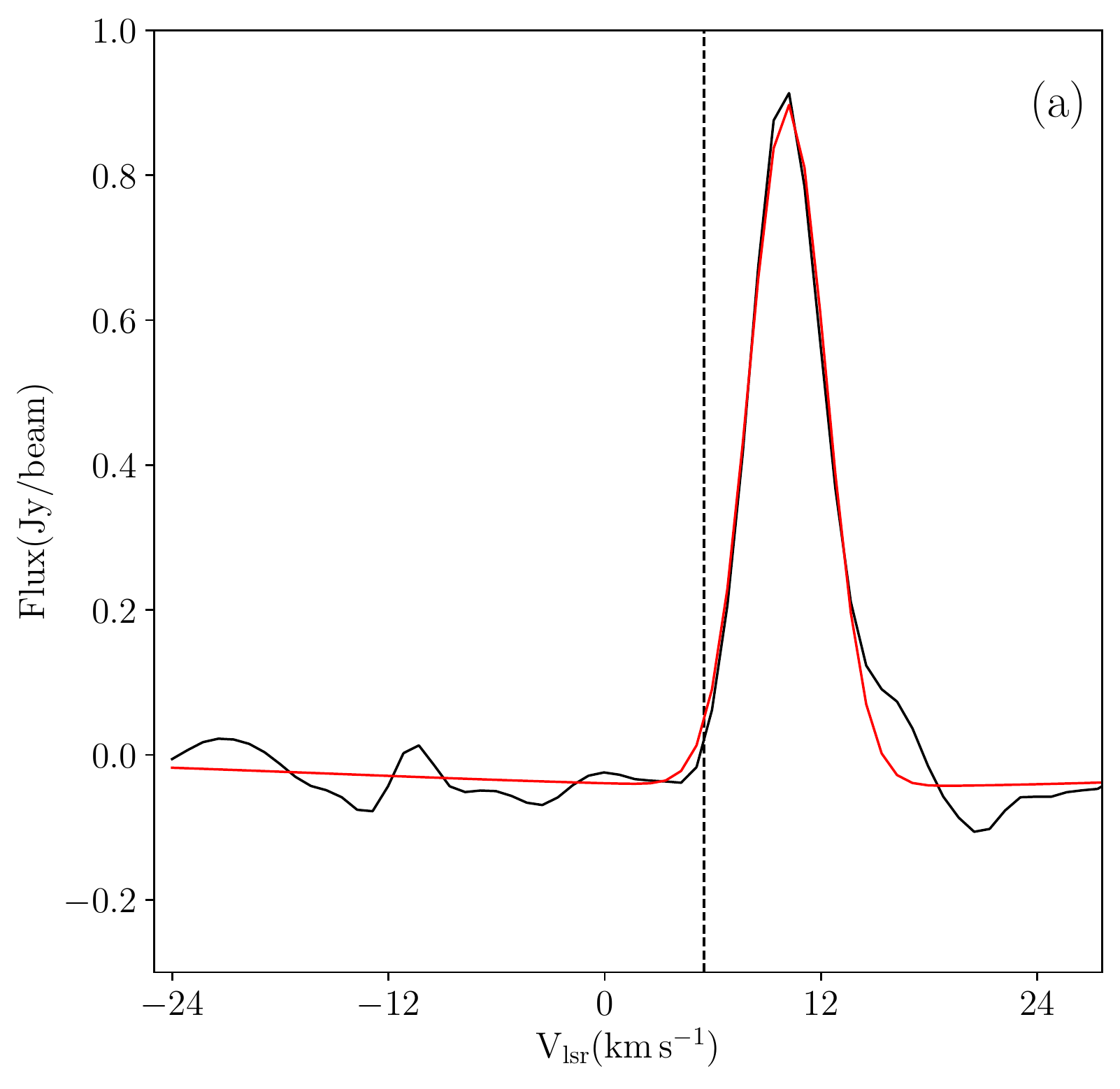}\includegraphics[scale=0.34]{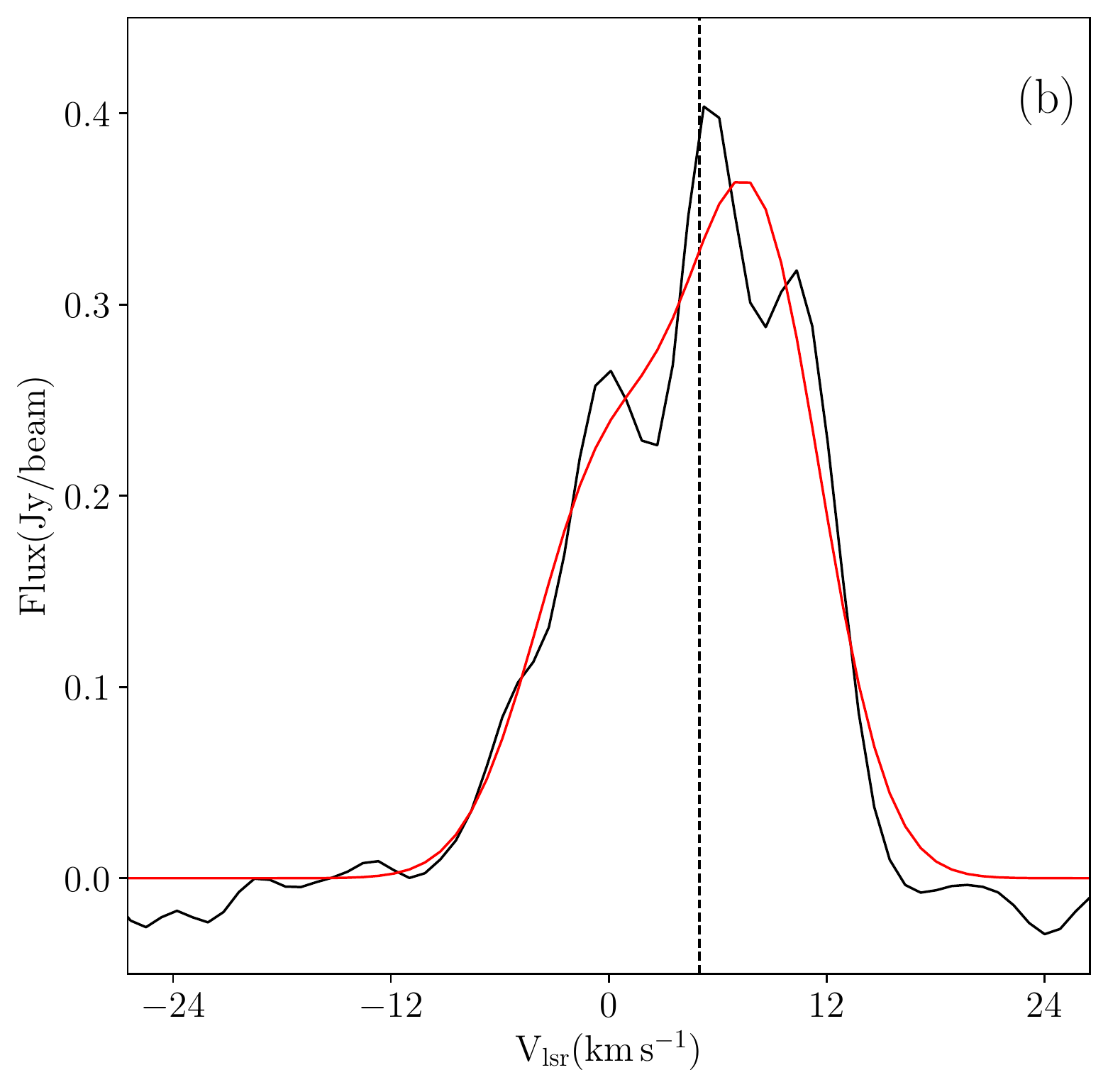}\includegraphics[scale=0.34]{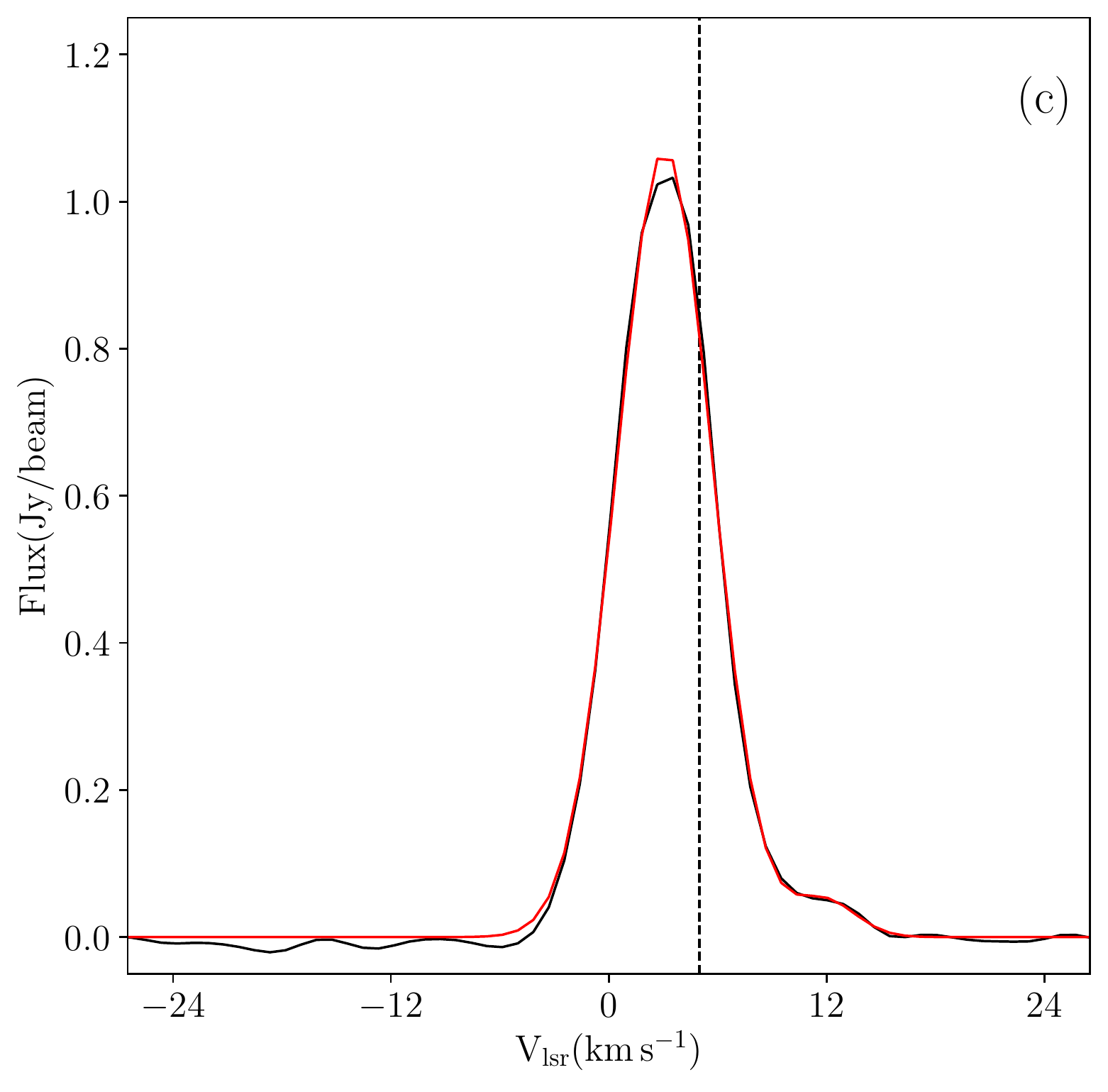}
\caption{Intensity profiles at the outflow axis of the position-velocity diagrams at a height $z=-80$ au in Figure \ref{fig:pvrot}. The red line shows the best Gaussian fit to the intensity profile of (a) $^{29}$SiO (J=8--7) $\nu=0$,
(b) SiS (J=19--18) $\nu=0$,
and (c) $^{28}$SiO (J=8--7) $\nu=1$. For reference, the dashed lines indicate the $V_{LSR}$ velocity.}
\label{fig:vexpapp}
\end{figure*}

\begin{figure*}[t!]
\centering
\includegraphics[scale=0.34]{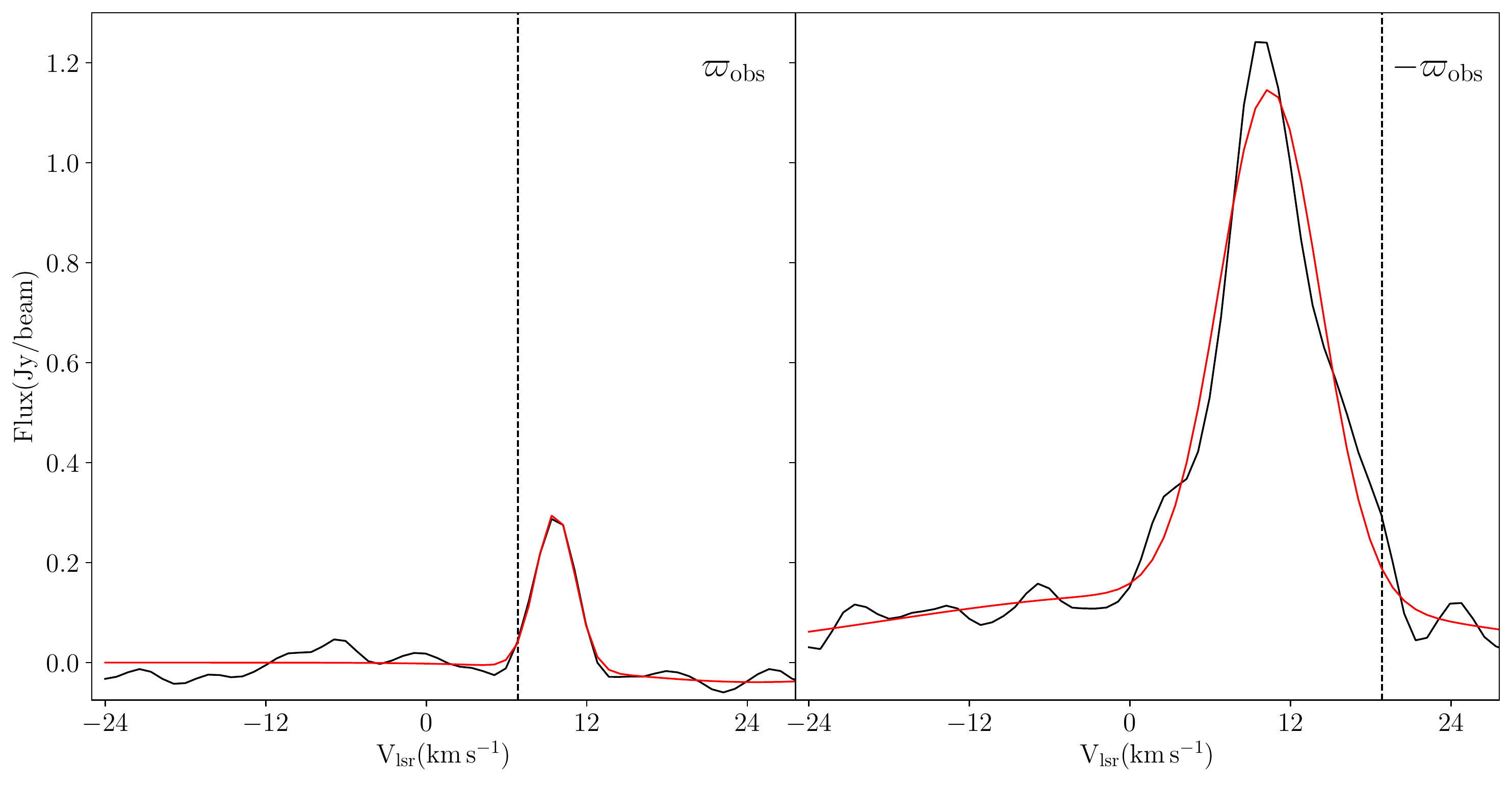}
\includegraphics[scale=0.34]{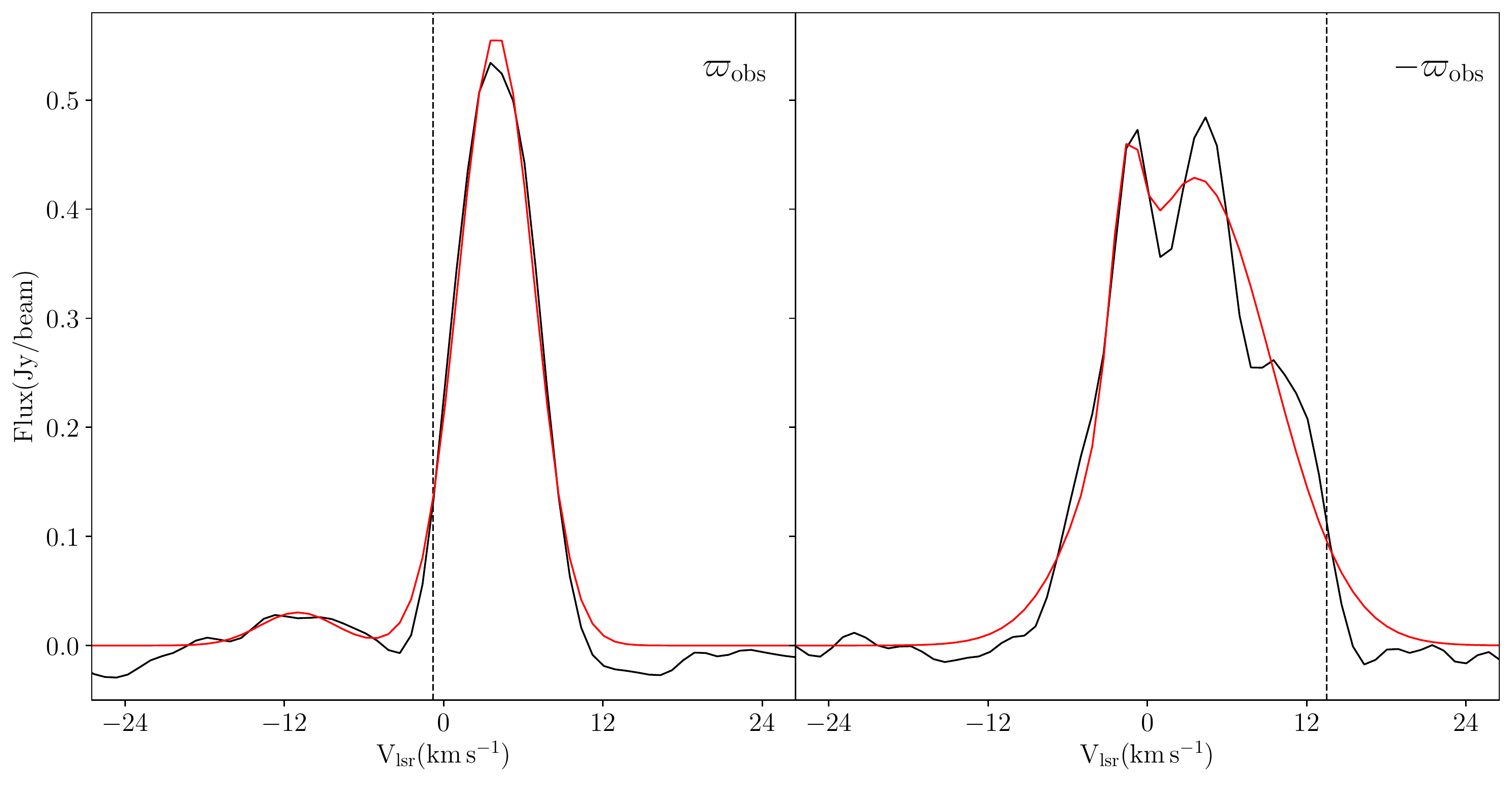}
\includegraphics[scale=0.34]{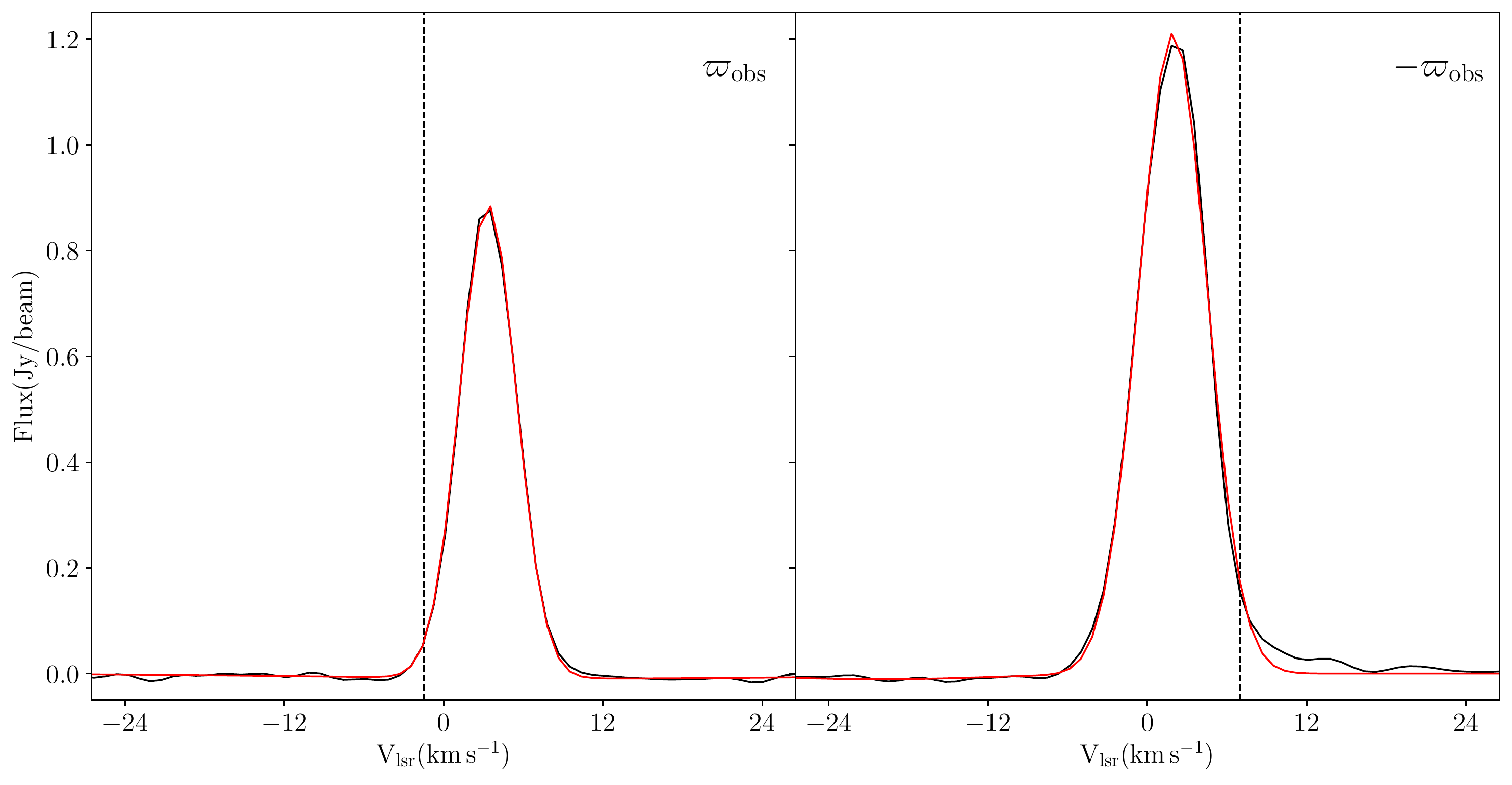}
\caption{Intensity profiles at the position of the cylindrical radii $\varpi_{\rm obs}$ (left panels) and $-\varpi_{\rm obs}$ (right panels) in 
 Figure \ref{fig:pvrot} at a height $z=-80$ au. The red line shows the best Gaussian fits to the intensity profiles 
 of $^{29}$SiO (J=8--7) $\nu=0$ (upper panels),  SiS (J=19--18) $\nu=0$ (middle panels), and $^{28}$SiO (J=8--7) $\nu=1$
 (lower panels).}
\label{fig:vrotapp}
\end{figure*}

The position-velocity diagrams in Figures \ref{fig:pv29sio}--\ref{fig:pvsio} were analyzed to derive the physical parameters:  the cylindrical radius $\varpi_{\rm obs}$, the expansion velocity $v_{\rm exp}$, and the rotation velocity $v_{\rm rot}$, as a function of the height $z$. These  properties were compared with the physical properties of the thin shell model of LV19.

Figure \ref{fig:routapp} shows the intensity profiles  at $V_{\rm LSR}=5$ km s$^{-1}$ as a function of the distance to the outflow axis at a height $z=-80$ au for the molecular lines $^{29}$SiO (J=8--7) $\nu=0$ (panel a), SiS (J=19--18) $\nu=0$ (panel b), and $^{28}$SiO (J=8--7) $\nu=1$ (panel c).  These panels also show a Gaussian fit to the intensity profiles (red solid lines). The cylindrical radius $\varpi_{\rm obs}$ 
is the width of the Gaussian profile
and the error if given by the Gaussian fit. In panel (b), the three peaks correspond to the emission from three shells. For our measurements, we only consider the two most prominent peaks. 
 The cylindrical radius of the shell model is the projection of the spherical radius $R_s$ at a given height, $\varpi_{\rm model}=R_s \sin\theta$, where $\theta=\cos^{-1}(z/R_s)$.

Figure \ref{fig:vexpapp} shows the intensity profiles at a height $z=-80$ au  at the outflow axis 
(angular offset =0 au in Figure \ref{fig:pvrot})
as a function of velocity for the three molecular lines, $^{29}$SiO (J=8--7) $\nu=0$ (panel a), SiS (J=19--18) $\nu=0$ (panel b), and $^{28}$SiO (J=8--7) $\nu=1$ (panel c).  The expansion
velocity is calculated at the outflow axis as $v_{\rm exp} = (v_+ - v_-)/2$, where $v_\pm$ are the radial 
velocities corresponding to the width of the Gaussian profile. 
 The axial velocity $v_{\rm z}$ is calculated as $v_{\rm z} = (v_+ + v_-)/2$. 
The errors are given by the Gaussian fit.  
In the case of the anisotropic stellar wind model, for a given inclination angle $i$, one calculates $v_\pm$ as the projection along the line of sight of the velocity of the two sides of the shell. 
The axial velocity is also corrected by the system velocity $V_{\rm LSR}$.

Figure \ref{fig:vrotapp} shows the intensity profiles as a function of the velocity at the cylindrical radii, $\varpi_{\rm obs}$ (left panels) and $-\varpi_{\rm obs}$ (right panels), shown as vertical dotted lines in Figure \ref{fig:pvrot},  for the three molecular lines, $^{29}$SiO (J=8--7) $\nu=0$ (upper panels), SiS (J=19--18) $\nu=0$ (middle panels), and $^{28}$SiO (J=8--7) $\nu=1$ (lower panels).  The red solid lines show the Gaussian fits, some of which require 2 Gaussians. 

The rotation velocity is given as the difference between the outer edges of widths of the intensity profiles at  $\pm \varpi_{\rm obs}$,
indicated by the dashed line in each panel 
(see also the inclined solid lines in Figure \ref{fig:pvrot}). The
error bars are given by  the Gaussian fit. For the model, we use the rotation velocity $v_\phi$.

 Figures \ref{fig:routapp}--\ref{fig:vrotapp} show, as an example, the analysis to obtain
the observed quantities $\varpi_{\rm obs}$, $v_{\rm exp}$, $v_{\rm z}$, and $v_{\rm rot}$ at  $z=-80$ au.
The same analysis is performed for each height $z$ in Figures \ref{fig:graphicsr} and \ref{fig:graphicsv}.



\listofchanges

\end{document}